\documentclass[12pt]{article}
\usepackage{a41}
\usepackage{cite}
\usepackage{color}
\usepackage{float}
\usepackage{amsmath}
\usepackage{amssymb}
\usepackage{amsxtra}
\usepackage{graphicx}
\usepackage{epstopdf}
\usepackage[rflt]{floatflt}
\usepackage{arrayjobx}
\allowdisplaybreaks[4]

\usepackage{mathtools}
\DeclarePairedDelimiter{\ceil}{\lceil}{\rceil}
\DeclarePairedDelimiter{\floor}{\lfloor}{\rfloor}

\newcommand{\gsim}{\raisebox{-0.07cm   }
{$\, \stackrel{>}{{\scriptstyle\sim}}\, $}}
\newcommand{\MS}{\overline{{\sf MS}}}
\newcommand{\Ahathat}{\hat{\hat{A}}}     
\newcommand{\ep}{\varepsilon}

\newcommand{\thep}{\frac{3}{2}\ep}

\newcommand{\N}{\nonumber}

\newcommand{\period}{\,.}
\newcommand{\comma}{\,,}

\newcommand{\Mvec}{{\bf M}}

\newcommand{\fracpart}[1]{\langle#1\rangle}
\newcommand{\intpart}[1]{\floor{#1}}

\newarray\ladderLabel
\readarray{ladderLabel}{1&2&3&4&5&6&7&8&9&10}

\newcounter{graphnr}
\newcommand{\HSums}{{\tt HarmonicSums}}
\newcommand{\CSsigma}{{\tt Sigma}}
\newcommand{\EvalMS}{{\tt EvaluateMultiSums}}

\newcommand{\NQGRAF}{{\tt QGRAF}}

\title{The 2-Loop Heavy Quark Corrections to Charged Current DIS}
\author{}

\usepackage[implicit=true]{hyperref}

\begin{document}
\setlength{\baselineskip}{0.515cm}
\sloppy
\thispagestyle{empty}
\begin{flushleft}
DESY 14--018
\hfill 
\\
DO-TH 14/04\\
MITP/14-019\\
SFB/CPP-14-24\\   
LPN 14-069\\
Higgstools 14-006\\       
May 2014\\
\end{flushleft}

\mbox{}
\vspace*{\fill}
\begin{center}
{\LARGE\bf The \boldmath $O(\alpha_s^3 T_F^2)$ Contributions to the Gluonic}

\vspace*{2mm}
{\LARGE\bf \boldmath Operator Matrix Element}

\vspace*{2mm}

\vspace{4cm}  
\large
J.~Ablinger$^a$,
J.~Bl\"umlein$^b$,
A.~De Freitas$^b$,
A.~Hasselhuhn$^{a,b}$,
A.~von~Manteuffel$^c$,
M.~Round$^{a,b}$, and
C.~Schneider$^a$

\vspace{1.5cm}
\normalsize
{\it $^a$~Research Institute for Symbolic Computation (RISC),\\
                          Johannes Kepler University, Altenbergerstra\ss{}e 69,
                          A--4040, Linz, Austria}\\

\vspace*{3mm}
{\it  $^b$ Deutsches Elektronen--Synchrotron, DESY,}\\
{\it  Platanenallee 6, D-15738 Zeuthen, Germany}

\vspace*{3mm}
{\it  $^c$ PRISMA Cluster of Excellence, Institute of Physics, J. Gutenberg University,}\\
{\it D-55099 Mainz, Germany.}

\end{center}
\normalsize
\vspace{\fill}
\begin{abstract}
\noindent 
The $O(\alpha_s^3 T_F^2 C_F (C_A))$ contributions to the transition matrix element $A_{gg,Q}$ relevant for 
the variable flavor number scheme at 3--loop order are calculated. The corresponding graphs contain two massive 
fermion lines of equal mass leading to terms given by inverse binomially weighted sums beyond the usual harmonic 
sums. In $x$-space two root-valued letters contribute in the iterated integrals in addition to those forming the 
harmonic polylogarithms. We outline technical details needed in the calculation of graphs of this type, 
which are as well of importance in the case of two different internal massive lines.
\end{abstract}

\vspace*{\fill}
\noindent
\numberwithin{equation}{section}
\newpage
\section{Introduction}

\vspace*{1mm}
\noindent
The precision determinations of  the strong coupling constant $\alpha_s(M_Z^2)$ \cite{Bethke:2011tr} 
and the parton densities, cf. e.g. Refs.~\cite{PDF}, in deep-inelastic scattering require 
the knowledge of the heavy flavor corrections to 3--loop order. The heavy flavor corrections were 
calculated at NLO in semi-analytic form in~\cite{HEAV2}\footnote{A fast and precise numerical
implementation in Mellin space has been given in~\cite{Alekhin:2003ev}.}. To avoid 
contributions of higher twist, the analysis has to be restricted to large enough 
values of $Q^2$. It has been shown in~\cite{BMSN96} that for $Q^2 \gsim 10~m^2$, with 
$m$ the heavy quark mass, the heavy flavor contributions to the structure function 
$F_2(x,Q^2)$ are very precisely described using the asymptotic representation in 
which all power corrections $\propto (m^2/Q^2)^k, k \in \mathbb{N}_+$ are neglected. 
In this limit the heavy flavor Wilson coefficients can be calculated analytically. They 
are given by convolutions of massive operator matrix elements (OMEs) and the massless 
Wilson coefficients, cf.~Ref.~\cite{BMSN96,BBK09NPB}. The massless Wilson coefficients 
are known to 3-loop order~\cite{MVV2005}. In the past the asymptotic $O(\alpha_s^2)$ corrections 
were calculated in Refs.~\cite{BMSN96,BBK07NPB,BMSN98,BBK09PLB,Buza:1996xr,Bierenbaum:2007pn,BBKS08NPB} 
in the unpolarized and polarized case, including the $O(\alpha_s^2 \ep)$ contributions, and in 
\cite{Blumlein:2009rg} for transversity. The heavy flavor corrections for charged current 
reactions are available at 1-loop and in the asymptotic case at 2-loops~\cite{CC}. 

At 3-loop order, a series of moments has been calculated for all massive 
OMEs for $N = 2 ... 10 (14)$ contributing in the fixed and variable flavor scheme 
\cite{BBK09NPB}. All logarithmic terms to 3-loop order including the contributions to the constant
term due to renormalization have been computed in Ref.~\cite{LOGS}. The  3-loop heavy flavor 
corrections to $F_L(x,Q^2)$ in the asymptotic case were calculated in \cite{Blumlein:2006mh,LOGS}. 
First results for general values of $N$ have been obtained for all OMEs for the color factor $N_F T_F^2 C_{F,A}$ 
\cite{ABKSW11NPB,Blumlein:2012vq} and 3-loop ladder, Benz-, and $V$-topologies \cite{ABHKSW12,Ablinger:2014yaa}.

First $\alpha_s^3 T_F^2 C_{F,A}$-contributions at general $N$ were calculated for the flavor non-singlet and pure-singlet 
terms in \cite{Ablinger:2011pb} for two heavy quark lines carrying the same mass. Furthermore, 
the moments $N = 2,4,6$ in case of the OMEs contributing to the structure 
function $F_2(x,Q^2)$ with two different heavy quark masses were computed in
\cite{Ablinger:2011pb,Ablinger:2012qj}. In all the above cases the massive OMEs are 
calculated for external massless partons which are on-shell. Recently, the complete 3-loop OMEs $A_{gq}, 
A_{qq,Q}^{\rm NS}$ and $A_{Qq}^{\rm PS}$ and the associated Wilson coefficients in the asymptotic region
have been calculated in Refs.~\cite{Ablinger:2014lka,NSPS}. Also the case of massive 
on-shell external lines has been treated in \cite{Blumlein:2011mi} recently.

In the present paper we calculate the $O(\alpha_s^3 T_F^2 C_{F,A})$ corrections to the massive 
OME  $A_{gg,Q}$ with local operator insertions on the gluonic lines at general values of $N$.
This matrix element is of importance to establish the variable flavor number scheme (VFNS) at 3--loop order.
The terms of $O(\alpha_s^3 T_F^2 C_{F,A})$ derive from graphs with two internal massive fermion lines of equal mass.
Unlike the foregoing 3-loop results for massive OMEs at general values of $N$ 
\cite{Blumlein:2006mh,ABKSW11NPB,LOGS,Ablinger:2014lka,NSPS} new functions beyond the harmonic sums \cite{HSUM}
appear, which belong to the finite nested binomially weighted harmonic sums \cite{ABRS14}. Here they are of the 
type\footnote{Infinite binomial sums of this kind have been studied in Refs.~\cite{BINO}.}
\begin{eqnarray}
\frac{1}{4^N} \binom{2N}{N} \sum_{k=1}^N \frac{4^k S_{\vec{a}}(k)}{k^l \binom{2k}{k}}~,
\end{eqnarray}
which have been considered in \cite{Fleischer:1998nb} before. Here $S_{\vec{a}}(N)$ denotes the nested harmonic sum
\begin{eqnarray}
S_{b,\vec{a}}(N) = \sum_{k=1}^N \frac{({\rm sign}(b))^k}{k^{|b|}} S_{\vec{a}}(k),~~~~S_\emptyset = 
1,~~~b, a_i \in \mathbb{Z} \backslash \{0\}~.
\end{eqnarray}
More involved sums of this type contribute to 
the massive $V$-topologies, cf. Ref.~\cite{Ablinger:2014yaa}. For a larger class of diagrams the calculation of the 
corresponding graphs is performed using Mellin-Barnes representations and requires cyclotomic harmonic sums and 
polylogarithms in intermediary steps. The corresponding nested sums are then solved using the summation and 
representation techniques 
encoded in the packages {\tt Sigma} \cite{SIGMA}, {\tt HarmonicSums} 
\cite{Harmonicsums,Ablinger:2011te,Ablinger:2013cf}, {\tt EvaluateMultiSums}, {\tt SumProduction} \cite{EMSSP}, 
and {\tt RhoSum}~\cite{RHOSUM}. For a few Feynman diagrams, it proved to be efficient to calculate them using 
integration-by-parts \cite{IBP}. The corresponding master integrals were computed applying systems of linear differential 
equations.

The paper is organized as follows. In Section~\ref{sec:2} we discuss the structure of the gluonic operator matrix 
element. At $O(\alpha_s^3 T_F^2 C_{F,A})$ 39 Feynman diagrams contribute. The calculation methods to obtain the
result at general values of the Mellin variable $N$ are outlined in Section~\ref{sec:3} in detail. In 
Section~\ref{sec:4} we present the results for the OME and also obtain the contributions $\propto T_F^2 C_{F,A}$
to the gluonic 3-loop anomalous dimension $\gamma_{gg}$. Section~\ref{sec:5} contains the conclusions.
In the Appendix we present the results for a series of scalar integrals which emerge in the present calculation.  
\section{The Operator Matrix Element}
\label{sec:2}

\vspace*{1mm}
\noindent

\vspace{1mm}
\noindent
The massive operator matrix element $A_{gg,Q}$ is the expectation value $\langle g|O_g|g\rangle$,  
of the gluonic operator  
\begin{eqnarray}
  O_{g,\mu_1,...,\mu_N}
  = 2 i^{N-2} \mathbf{S}\text{Sp}
  [F_{\mu_1\alpha} D_{\mu_2} ... D_{\mu_{N-1}}F^{\alpha}_{\phantom{\alpha}\mu_N}]
  - \text{trace terms}
\label{eq:OP}
\end{eqnarray}
between massless on-shell external gluon states. We will work in $R_\xi$--gauge. Therefore also the corresponding 
ghost graphs have to be considered. In Eq.~(\ref{eq:OP}), $\mathbf{S}$ and
$\text{Sp}$ denote the symmetrization of the Lorentz indices and color trace, respectively; $F_{\mu\nu}$ is the field
strength tensor of QCD and $D_{\alpha}$ denotes the covariant derivative.
The OME has been calculated to $O(\alpha_s^2)$ in \cite{BMSN98} and including also terms linear in $\ep$ in 
\cite{BBK09PLB} correcting the previous result.

The renormalized expression of $A_{gg,Q}$ to $O(\alpha_s^3)$ was derived in 
\cite{BBK09NPB} and the contributions to $O(\alpha_s^3 T_F^2 N_F C_{F,A})$ were calculated in
\cite{Blumlein:2012vq}. 
The OME $A_{gg,Q}$ obeys the expansion
\begin{eqnarray}
A_{gg,Q}(N,a_s) = \tfrac{1}{2} [1 + (-1)^N] \left\{ 1 + \sum_{k=1}^\infty a_s^k A_{gg,Q}^{(k)}(N)\right\},
\end{eqnarray}
with $a_s(\mu^2) = \alpha_s(\mu^2)/(4\pi)$.
In the $\overline{\rm MS}$ scheme with the heavy quark mass $m$ on-shell\footnote{For the representation in 
the $\overline{\rm MS}$-scheme for the heavy quark mass, see Section~\ref{sec:4}.}
it is given by
\begin{eqnarray}
\label{eqAggQ}
  A_{gg,Q}^{(3), \MS}&=&
                    \frac{1}{48}\Biggl\{
                            \gamma_{gq}^{(0)}\hat{\gamma}_{qg}^{(0)}
                                \Bigl(
                                        \gamma_{qq}^{(0)}
                                       -\gamma_{gg}^{(0)}
                                       -6\beta_0
                                       -4n_f\beta_{0,Q}
                                       -10\beta_{0,Q}
                                \Bigr)
                           -4
                                \Bigl(
                                        \gamma_{gg}^{(0)}\Bigl[
                                            2\beta_0
                                           +7\beta_{0,Q}
                                                         \Bigr]
\N
\\ 
&&
                                       +4\beta_0^2
                                       +14\beta_{0,Q}\beta_0
                                       +12\beta_{0,Q}^2
                                \Bigr)\beta_{0,Q}
                     \Biggr\}
                     \ln^3 \Bigl(\frac{m^2}{\mu^2}\Bigr)
                    +\frac{1}{8}\Biggl\{
                            \hat{\gamma}_{qg}^{(0)}
                                \Bigl(
                                        \gamma_{gq}^{(1)}
                                       +(1-n_f)\hat{\gamma}_{gq}^{(1)}
                                \Bigr)
\N\\ &&
                           +\gamma_{gq}^{(0)}\hat{\gamma}_{qg}^{(1)}
                           +4\gamma_{gg}^{(1)}\beta_{0,Q}
                           -4\hat{\gamma}_{gg}^{(1)}[\beta_0+2\beta_{0,Q}]
                           +4[\beta_1+\beta_{1,Q}]\beta_{0,Q}
\N\\ &&
                           +2\gamma_{gg}^{(0)}\beta_{1,Q}
                     \Biggr\}
                     \ln^2 \Bigl(\frac{m^2}{\mu^2}\Bigr)
                    +\frac{1}{16}\Biggl\{
                            8\hat{\gamma}_{gg}^{(2)}
                           -8n_fa_{gq,Q}^{(2)}\hat{\gamma}_{qg}^{(0)}
                           -16a_{gg,Q}^{(2)}(2\beta_0+3\beta_{0,Q})
\N\\ &&
                           +8\gamma_{gq}^{(0)}a_{Qg}^{(2)}
                           +8\gamma_{gg}^{(0)}\beta_{1,Q}^{(1)}
                   +\gamma_{gq}^{(0)}\hat{\gamma}_{qg}^{(0)}\zeta_2
                                \Bigl(
                                        \gamma_{gg}^{(0)}
                                       -\gamma_{qq}^{(0)}
                                       +6\beta_0
                                       +4n_f\beta_{0,Q}
                                       +6\beta_{0,Q}
                                \Bigr)
\N\\ &&
                   +4\beta_{0,Q}\zeta_2
                                \Bigl( 
                                       \gamma_{gg}^{(0)}
                                      +2\beta_0
                                \Bigr)
                                \Bigl(
                                       2\beta_0
                                      +3\beta_{0,Q}
                                \Bigr)
                     \Biggr\}
                     \ln \Bigl(\frac{m^2}{\mu^2}\Bigr)
                   +2(2\beta_0+3\beta_{0,Q})\overline{a}_{gg,Q}^{(2)}\N
\N\\ &&
                   +n_f\hat{\gamma}_{qg}^{(0)}\overline{a}_{gq,Q}^{(2)}
                   -\gamma_{gq}^{(0)}\overline{a}_{Qg}^{(2)}
                   -\beta_{1,Q}^{(2)} \gamma_{gg}^{(0)}
                   +\frac{\gamma_{gq}^{(0)}\hat{\gamma}_{qg}^{(0)}\zeta_3}{48}
                                \Bigl(
                                        \gamma_{qq}^{(0)}
                                       -\gamma_{gg}^{(0)}
                                       -2[2n_f+1]\beta_{0,Q}
\N\\ &&
                                       -6\beta_0
                                \Bigr)
                   +\frac{\beta_{0,Q}\zeta_3}{12}
                                \Bigl(
                                        [\beta_{0,Q}-2\beta_0]\gamma_{gg}^{(0)}
                                       +2[\beta_0+6\beta_{0,Q}]\beta_{0,Q}
                                       -4\beta_0^2
                                \Bigr)
\N\\ &&
                   -\frac{\hat{\gamma}_{qg}^{(0)}\zeta_2}{16}
                                \Bigl(
                                        \gamma_{gq}^{(1)}
                                       +\hat{\gamma}_{gq}^{(1)}
                                \Bigr)
                   +\frac{\beta_{0,Q}\zeta_2}{8}
                                \Bigl(
                                        \hat{\gamma}_{gg}^{(1)}
                                      -2\gamma_{gg}^{(1)}
                                      -2\beta_1
                                      -2\beta_{1,Q}
                                \Bigr)
                           +\frac{\delta m_1^{(-1)}}{4}
                                \Bigl(
                                     8 a_{gg,Q}^{(2)}
\N\\ &&
                                    +24 \delta m_1^{(0)} \beta_{0,Q}
                                    +8 \delta m_1^{(1)} \beta_{0,Q} 
                                    +\zeta_2 \beta_{0,Q} \beta_0
                                    +9 \zeta_2 \beta_{0,Q}^2
                                \Bigr)
                           +\delta m_1^{(0)}
                                \Bigl(
                                     \beta_{0,Q} \delta m_1^{(0)}
                                    +\hat{\gamma}_{gg}^{(1)}
                                \Bigr)
\N\\ &&
                           +\delta m_1^{(1)}
                                \Bigl(
                                     \hat{\gamma}_{qg}^{(0)} \gamma_{gq}^{(0)}
                                    +2 \beta_{0,Q} \gamma_{gg}^{(0)}
                                    +4 \beta_{0,Q} \beta_0
                                    +8 \beta_{0,Q}^2
                                \Bigr)
                           -2 \delta m_2^{(0)} \beta_{0,Q}
                 +a_{gg,Q}^{(3)}~. 
        \label{Agg3QMSren}
   \end{eqnarray}
Here $\delta m_i^{(k)}$ are expansion coefficients of the renormalization constants for the  mass,
$\beta_i,\beta_{i,Q}$ are coefficients of the $\beta$-functions (including mass 
effects), $\zeta_k$ is the Riemann--$\zeta$ function with $k \in \mathbb{N} \backslash 
\{0,1\}$, $a^{(2)}_{ij}, \overline{a}^{(2)}_{ij}$
are two loop contributions to order $\ep^0$ and $\ep^1$, respectively,
and $\gamma_{ij},\hat{\gamma}_{ij}$ are the anomalous dimensions. Quantities with
a hat in Eq.~(\ref{Agg3QMSren}) are defined by
\begin{eqnarray}
  \hat{f} = f(n_f+1) - f(n_f), 
\end{eqnarray}
see Ref.~\cite{BBK09NPB}.
The unrenormalized OME $\Ahathat_{gg,Q}^{(3)}$ also receives contributions from the 
vacuum polarization insertions on the external lines
\begin{eqnarray}
\hat{\Pi}^{ab}_{\mu\nu}(p^2,\hat{m}^2,\mu^2,\hat{a}_s^2) 
 &=& i\delta^{ab}\left[-g_{\mu\nu} p^2 + p_\mu p_\nu\right]
  \sum_{k-1}^{\infty}\hat{a}_s^k \hat{\Pi}^{(k)}(p^2,\hat{m}^2,\mu^2)
\\
  \hat{\Pi}^{(k)} &\equiv& \hat{\Pi}^{(k)}(0,\hat{m}^2,\mu^2)
\end{eqnarray}
such that
\begin{eqnarray}
   \Ahathat_{gg,Q}^{(3)}
   &=&
            \Ahathat_{gg,Q}^{(3), \text{1PI}}
           -\hat{\Pi}^{(3)}
           -\Ahathat_{gg,Q}^{(2), \text{1PI}}
            \hat{\Pi}^{(1)}
           -2\Ahathat_{gg,Q}^{(1)}
            \hat{\Pi}^{(2)}
           +\Ahathat_{gg,Q}^{(1)}
            \hat{\Pi}^{(1)}
            \hat{\Pi}^{(1)}
\\
  &\equiv&
   \frac{a_{gg,Q}^{(3,0)}}{\ep^3}
  +\frac{a_{gg,Q}^{(3,1)}}{\ep^2}
  +\frac{a_{gg,Q}^{(3,2)}}{\ep}
  +a_{gg,Q}^{(3)}~.
\label{unren}
\end{eqnarray}
All contributions to Eq.~(\ref{eqAggQ}) but the constant terms 
$a_{ij,Q}^{(3)}$ are 
known~\cite{BMSN96,BBK07NPB,BBKS08NPB,BMSN98,BBK09PLB,Vogt:2004mw}. In particular, all the
logarithmic contributions have already been obtained for general values of the Mellin variable $N$
\cite{Bierenbaum:2010jp,LOGS}.

In the following we calculate the $O(a_s^3 T_F^2 C_{F,A})$-contributions to the 
massive gluonic OME. Before  presenting the results, we give a detailed outline of the calculation methods used.
\section{The Methods of Calculation}
\label{sec:3}

\vspace*{1mm}
\noindent
The $T_F^2 C_{F,A}$-contributions to $A_{gg,Q}^{(3)}$ are given by Feynman graphs with external on-shell gluons (ghosts), 
a local operator insertion on gluon lines and vertices, and two closed massive quark lines of the same 
mass $m$. A calculation along the lines of Refs.~\cite{ABKSW11NPB,ABHKSW12} leads to infinite series which diverge 
polynomially with degree $N$. The way to cure this issue will be to separate the variable $N$ from the infinite 
series by leaving one integral unintegrated. This last integral will then be solved after summation in the space 
of cyclotomic harmonic polylogarithms \cite{Ablinger:2011te}. Most of the graphs have been calculated in this way. 
For a few graphs, we have applied integration by parts and differential equations, see~Section~\ref{sec:3.n}. 
Throughout the calculation, the results at general values of $N$ are mutually compared to the corresponding moments
calculated using {\tt MATAD} \cite{Steinhauser:2000ry}.
\subsection{Feynman Parameterization}

\vspace*{1mm}
\noindent
The list of graphs was generated with \NQGRAF{} \cite{Nogueira:1991ex} and
written as momentum integrals using the  Feynman rules of
\cite{BBK09NPB,Klein:2009ig}\footnote{For the scalar Feynman rules
used for the calculation of scalar prototype graphs, see
\cite{Hasselhuhn:2013swa}.}.  The color-algebra was performed using the code
{\tt Color} \cite{vanRitbergen:1998pn}.
The momenta were integrated at the cost
of introducing a Feynman parameterization, treating each independent
loop separately and introducing for each one of them a family of Feynman
parameters.  This makes each diagram a linear combination of integrals
of the form
\begin{align}
 \int_{[0,1]^{n}} dx_1\;\dots dx_n
 \left(\prod_{\text{families } f}\delta^f\right)
 \underbrace{x_1^{\nu_1-1} \dots x_n^{\nu_n-1}}_{\text{monomial
prefactor}}
 \underbrace{
 \prod_{i=1}^{n} x_i^{\alpha_i} (1-x_i)^{\beta_i}}_{\text{non-monomial prefactor}}
 \frac{{\overbrace{P_O(x_1,\dots,x_n;N)}^{\text{operator
polynomial}} } }
  {{\underbrace{[P_D(x_1,\dots,x_n)]}_{\text{\parbox[t]{6em}{denominator
\\[-0.5em]
polynomial}} }}{}^\gamma }
\comma
\end{align}
where for each Feynman parameter family $f$ we used the short-hand notation
\begin{align}
 \delta^f \equiv \delta\left(1-\sum_{x\in f}x\right)
\comma
\end{align}
and $\nu_i$ are integers denoting the propagator powers. The exponents
$\alpha_i,\beta_i,\gamma$ are of the form $(a+b\ep/2)$ with $a,b
\in\mathbb{Z}$, and $N$ is the Mellin variable.  The operator
polynomial is not strictly a polynomial, but in all following cases
the $\delta$-distributions and Heaviside functions being present in addition 
can be removed in such a way that the misnomer is corrected, and the operator polynomial
is indeed a polynomial of maximum degree $N \in \mathbb{N}$. 

The $\delta$-distributions can be integrated using the relations
\begin{align}
 \int_0^1 dx\; \delta(1-x-Y) f(x) ={}& \theta(Y) \theta(1-Y)
f(1-Y)\comma
\end{align}
and
\begin{align}
\label{eq:MapTheta}
 \int_0^1 dx\; \theta(1-x-Y) f(x) 
 ={}& 
 \int_0^1 dx\; \theta(1-Y) (1-Y) f(x(1-Y))
 \comma
\end{align}
where $Y$ is either a sum of Feynman parameters or a single one.
The Heaviside $\theta$-function is defined as
\begin{align}
\label{eq:defHeaviside}
  \theta(x) 
  = 
  \left\{
    \begin{array}{ll} 1 ,& x\ge 0 \\ 0, & x<0 \end{array}
  \right.
\period
\end{align}
These relations are applied in such a way as to keep the operator
polynomial as simple as possible. It is indeed possible in all
following cases, to map the operator polynomial into one single
Feynman parameter, if one uses the following trick: In some cases it is
useful to reconstruct a $\delta$-distribution by
\begin{align}
\label{eq:deltaReconst}
  \theta(X) \theta\left(1-X\right) f\left(1-X\right)
  ={}&
  \int_0^1 dy
  \delta\left(1-X-y\right) f\left(1-X\right)
\N\\
  ={}&
  \int_0^1 dy
  \delta\left(1-X-y\right) f(y)
\comma
\end{align}
where $X$ represents a sum of Feynman parameters.  Of course the order
for the elimination of the Feynman parameters from the
$\theta$-functions has to be chosen such that the left hand side of
the above equation matches. In this way, an argument $(1-X)$
consisting of several Feynman parameters is exchanged for only one
Feynman parameter. This trick is equivalent to a set of coordinate
transformations mentioned in \cite{Hamberg91} and also used in the
calculation of the 2-loop OMEs in \cite{Bierenbaum:2007dm,Bierenbaum:2007qe,
BBKS08NPB,BBK09PLB,ABKSW11NPB,Blumlein:2012vq}.
The above trick has the advantage of giving a clear
guideline for how to simplify the polynomial in the $N$-bracket of the
Feynman integrals under consideration.

It is worth noting that there are two Feynman parameters, which only
occur in the monomial prefactors of the integrand as well as in the
operator polynomial. These are due to the fact that the incoming and
outgoing momenta are massless. The integral over these Feynman
parameters can thus be performed easily, giving simpler $N$-brackets.

The above methods are applied in order to avoid the proliferation of
$N$. In fact, in all diagrams one can achieve that $N$ only
occurs in the exponent of one of the Feynman parameters, allowing to
effectively decouple $N$ from the solution of infinite sums.  This
property of the calculation is of crucial importance, and also
carries over to the case of two lines of unequal masses which,
however, will be the subject of a future publication.
\subsection{Mellin-Barnes Representation}

\vspace*{1mm}
\noindent
The remaining parameters still occur in the denominator polynomial.
It has the form $(A+B)$ where $A$ and $B$ are products of
elements $x_i$ or $(1-x_i)$, for Feynman parameters $x_i$.  Only in
the cases of graphs with a massive line that runs through four edges
of the graph, e.g.\  graphs \ref{gra:rb2} and \ref{gra:rb} in Appendix
\ref{app:Results}, a factor $(1-x(1-y))$ in either $A$ or $B$ occurs.
A Mellin-Barnes (MB) integral \cite{Mellin1895, Barnes1908} is then
introduced by the substitution, see e.g. \cite{WWT,Smirnov2006},
\begin{align}
 (A+B)^{-\gamma}
 =
 \frac{1}{\Gamma(\gamma)}
 \frac{1}{2\pi i}
 \int_{-i\infty}^{i\infty} d\xi
 \Gamma(-\xi)\Gamma(\gamma+\xi)
 \frac{A^{\xi}}{B^{\gamma+\xi}}~.
\end{align}
This procedure is equivalent to splitting the mass-term off the
propagator-like part that occurs in the Feynman parameter
representation of a massive vacuum polarization diagram, before
proceeding with successive parameterization and momentum integration. 

In the cases that the products $A,B$ from above factorize completely,
all integrals can be performed in terms of Euler's Beta-functions. In the
remaining two cases, in which a factor $(1-x(1-y))$ remains, the
integrals represent a generalized hypergeometric function ${}_3F_2$ \cite{HYP,SLATER},
which in the scalar diagrams is already given in a form such that it
reduces to a ratio of $\Gamma$-functions. In the corresponding
physical cases, these functions lead to double sums, which can be
constructed such that they converge, still keeping $N$ separated from
the sums in the way described above.

At this point in all the diagrams only one Beta-function remains that
contains both $N$ and $\xi$.  This function is rewritten in terms of a
Feynman parameter integral, i.e. for corresponding $\alpha$ and
$\beta$
\begin{align}
\label{eq:BetaFP}
 B(N+\xi+\alpha, -\xi + \beta)
 =
 \int_0^1 dx\;
 x^{N+\xi+\alpha-1}
 (1-x)^{\beta - \xi -1}
\period
\end{align}
The reason is that the contour of the Mellin-Barnes integral cannot
be closed to a single side.  One can see this from two representations.
On the one hand, the Beta-function which contains $N$ and $\xi$ has the
form
\begin{align}
 B(N+\xi+\alpha, -\xi + \beta)
 =
 \frac{\Gamma(N+\xi+\alpha) \Gamma(-\xi + \beta)}
 {\Gamma(N+\alpha + \beta)}
\comma
\end{align}
so that the denominator drops out of the MB-integral. Hence, if the
contour is closed to one side and written as the sum of residues,
then, due to its convergence condition, for every set of values of the
propagator powers there is an $N_0$ so that for $N>N_0$ the sum is
divergent. 

On the other hand, if the Beta-function is written as a Feynman
parameter integral over $x$, then the factor
\begin{align}
  \left(\frac{1-x}{x}\right)^{\xi},
\end{align}
occurs in the integrand. Here a
distinction is necessary between values $x<\frac{1}{2}$ for which the
contour may be closed towards $\xi\rightarrow \infty$, and values
$x>\frac{1}{2}$ for which $\xi\rightarrow -\infty$ is the convergent
choice.  For simplicity, we change the order of the $\xi$-integration
such that the contour can be closed to the right in all cases.

After that the quantity raised to the power $\xi$ is mapped onto a
single integration variable $T$
\begin{align}
\label{eq:mapHalfToFull}
 T \equiv \frac{x}{1-x} \in [0,1] \quad 
 \Leftrightarrow &
 \quad x \equiv \frac{T}{1+T} \in \left[0,\frac{1}{2}\right]
 \comma
\N\\
 T \equiv \frac{1-x}{x} \in [0,1] \quad 
 \Leftrightarrow &
 \quad x \equiv \frac{1}{1+T} \in \left[\frac{1}{2},1\right]
 \comma
\N\\
 \text{with }~~~~~
 dx ={}& \frac{1}{(1+T)^2} dT
 \period
\end{align}
Now it is obvious that all contours have to be closed to the right
before applying the residue theorem.

It is worthwhile having a look onto convergence issues of the
procedure described so far.  First, the Mellin-Barnes integral is
introduced in the integrand of the multiple Feynman parameter
integral. Employing the nomenclature of \cite{Smirnov2006}, the
contour follows the usual requirement that left-poles (poles of
functions $\Gamma(\cdots-z)$) are to the left of the contour, and
right-poles (poles of $\Gamma(\cdots+z)$) are to the right of the
contour. If left- and right-poles are interleaved on the real axis,
the contour winds around them separating the two types of poles.

Of course, contours of the above kind can only be found, if the
right-poles are separated from left-poles.  In cases where this is not
obviously the case, we enforce such a separation by introducing a
regularization parameter in a consistent manner throughout the Feynman 
diagram.  So it
is most convenient to keep symbolic propagator powers from the
beginning, and to use substitutions of these symbolic quantities for
the introduction of regulators.  We will see later at which point the
expansion into a Laurent series in these parameters can be performed
most conveniently. 

The classical procedure for calculating Mellin-Barnes integrals in
particle physics proceeds by deforming the contour and subtracting
a finite number of residues, such that the remaining contour integral
represents a regular function in $\ep$  \cite{Smirnov2006, Tausk:1999vh,
Czakon:2005rk, Gluza:2007rt}.  In that case, the expansion can be performed
on the integrand level, which simplifies the integrand such that
Barnes lemmas are applicable.  However, since factors of $T^\xi$ occur
in the arguments of the contour integrals, cf. Eqs.~\ (\ref{eq:BetaFP}, \ref{eq:mapHalfToFull}), no Barnes lemmas 
\cite{Barnes1908}
can be applied\footnote{For a list of corollaries see
\cite{Smirnov2006}. For an automated use of Barnes' lemmas see the {\tt Mathematica} package {\tt barnesroutines} 
\cite{KOSOW}.}.  

In the present calculation, it appears more suitable to write
down the sums of residues and generate the necessary simplifications
and algebraic relations by symbolic summation methods implemented in
the package \CSsigma{} \cite{SIGMA}, equipped with
suitable limit procedures for infinite sums.

When residues are calculated and the corresponding sums are written down,
one has to perform a Laurent expansion in the regularization
parameters.  Here it is important to observe the singularity
structure.

One therefore brings the $\Gamma$-function arguments to a standard
form, such that all of them are positive for vanishing regulators
\begin{align}
  \Gamma(x) 
  = 
   \theta(\intpart{x}-1)\Gamma(x) 
  +
   \theta(-\intpart{x})
   (-1)^{\intpart{x}+1}
   \frac{\Gamma(\fracpart{x})\Gamma(1-\fracpart{x})}
   {\Gamma(1-x)} 
   \period
\end{align}
Here $\fracpart{x}$ and $\intpart{x}$ represent the fractional and
integer parts of the variable $x$, respectively.  The regulators are
assumed to be small enough, such that they only contribute to the
fractional part. The Heaviside functions are removed by commuting them with summation
operators. This can be done using the following operator relations
\begin{align}
  \sum_{i=a}^{b} \theta(c + d \cdot i) 
  ={}& 
  \theta\left(\ceil*{-\frac{c}{d}} - a\right)
  \theta\left(b-\ceil*{-\frac{c}{d}}\right)
  \sum_{i=\ceil*{c/d}}^{b}
  +
  \theta\left(a-\ceil*{-\frac{c}{d}}-1\right)
  \sum_{i=a}^{b}
\comma
\N\\
  \sum_{i=a}^{b} \theta(c - d \cdot i)
  ={}& 
  \theta\left(\floor*{\frac{c}{d}} - a\right)
  \theta\left(b-\floor*{\frac{c}{d}} - 1\right)
  \sum_{i=a}^{\floor*{c/d}}
  +
  \theta\left(\floor*{\frac{c}{d}} - b\right)
  \sum_{i=a}^{b}
\period
\end{align}
Once the $\theta$-functions are free of any summation parameters,
they can be evaluated. Note that they are also free of the Mellin
variable $N$, since it had been separated from the sums by
construction.  

Once the $\Gamma$-functions have been reflected such that the integer
parts of their arguments are positive using the relation \cite{SLATER}
\begin{eqnarray}
\Gamma(-N + x) = (-1)^N \frac{\Gamma(x) \Gamma(1-x)}{\Gamma(N+1-x)},~~~~~x \in \mathbb{R}, N \in \mathbb{N},
\end{eqnarray}
their expansion in the
artificial regulators is straightforward.  

Yet one additional preparation is necessary for the expansion in the
dimensional regulator $\ep$, since the Feynman parameter integrals may
not be well defined in the Lebesgue sense for $0<\ep<1$, but rather as
an analytic continuation in $\ep \rightarrow 0$.  The expressions are of
the form
\begin{align}
 f(\ep) = \int_0^1 dx\; x^{\ep-a} g(x)
\comma
\end{align}
which only converges if $\ep>a-1$.
Nevertheless, using integration by parts, one can shift this integrand
such that it is integrable for $0 < \ep < 1$.  For the form above with
$a \ge 1$, the relation
\begin{align}
  \int_0^1 dx\; x^{\ep-a} g(x)
  =
  \frac{g(1)}{\ep-a+1}
  -
  \frac{1}{\ep-a+1}
  \int_0^1 dx\; x^{\ep-a+1} g'(x)
\end{align}
has to be iterated $(a-1)$-times. Here the function $g(x)$ must have
sufficiently many regular derivatives on $[0,1]$, which is indeed the
case for the integrals in question.  Then the integral represents a
regular function in $\ep$, the integrand is measurable for $0 \le \ep
<1$ and thus the Taylor expansion commutes with the integration.

Finally, the expansion of the sums in the dimensional regulator $\ep$ can be done using the package \EvalMS{}.  It also manages 
the call of \CSsigma{} routines and performs limits of many expressions. In particular, the package {\tt SumProduction} was
used to condense the huge expressions into a tractable number of compact but larege sums and automatically apply the
summation technologies to obtain the final results.

The result of expansion and summation yields an expression which
still depends on one integration variable $T$, and which contains
S-sums \cite{Moch:2001zr,Ablinger:2013cf} and cyclotomic S-sums \cite{Ablinger:2011te} of this variable. They can be 
converted
into (cyclotomic) harmonic polylogarithms (HPL) \cite{Ablinger:2011te}, e.g.
\begin{align}
\label{eq:CS11relH}
S_{(2,1,1),(2,1,1)}(-T,1;\infty ) =
-\frac{H_{(4,0)}\left(\sqrt{T}\right)}{\sqrt{T}}+\frac{H_{0,(4,0)}\left(\sqrt{T}\right)}{\sqrt{T}}
-\frac{H_{(4,1),(4,0)}\left(\sqrt{T}\right)}{\sqrt{T}}~.
\end{align}
The conversions to iterated integrals are performed using ideas of
Ref.~\cite{Ablinger:2011te} and applying automated routines of the
package \HSums{} \cite{Harmonicsums,Ablinger:2011te,Ablinger:2013cf}.

The conversion returns iterated integrals evaluated at 1,
but with letters that depend on the remaining integration variable.
They will be denoted by
\begin{align}
\label{eq:scaleLetter}
  f_{[\alpha,y]}(x) := f_{\alpha}(xy)
\comma
\end{align}
where $f_{\alpha}$ is a letter from a cyclotomic alphabet
\begin{align}
  \Biggl\{
    f_0(x)=\frac{1}{x},
    f_1(x)=\frac{1}{1-x},
    f_{-1}(x)=\frac{1}{1+x},
    f_{(4,0)}(x)=\frac{1}{1+x^2},
    f_{(4,1)}(x)=\frac{x}{1+x^2}
  \Biggr\}
\period
\end{align}
Therefore a procedure is needed that maps the class of iterated
integrals appearing here onto (cyclotomic) HPLs with the integration
variable in the argument.  There is such a procedure which was used
for deriving properties of two-dimensional HPLs \cite{Gehrmann:2001jv} and in
the method of hyperlogarithms \cite{Brown:2008um,ABHKSW12,Ablinger:2014yaa}.
It makes use of the fact that differentiation of a certain type of
iterated integrals with respect to variables appearing in the index
leads to a drop in the weight of the function, e.g.  
\begin{align}
 \frac{\partial}{\partial x}
 H_{-x,-1}(1)
 =
 \frac{\partial}{\partial x}
 \int_0^1\frac{dy}{x+y}\int_0^y\frac{dz}{1+z}
 =
 -\frac{H_{-x}(1)}{1-x}
 -\frac{2 \ln(2)}{x^2-1}
 \comma
 \quad
 x>0
 \period
\end{align}
In this way the problem can be traced back to properties of rational
functions, solving the problem recursively at a lower weight and
integrating again over $x$, where in each recursive call a constant has
to be determined.

However, in the case of letters containing polynomials of degree 2 or
more, this procedure is not applicable directly, since in general the weight does not drop
due to differentiation, e.g.
\begin{align}
\frac{\partial}{\partial x} H_{[(4.0),x],-1}(1) - 
\frac{1}{x} H_{[(4,0),x],-1}(1) - \frac{x}{x^2+1} H_{[(4,1),x]}(1) - \frac{1}{x(x^2+1)} H_{[(4,0),x]}(1)
+ \frac{2 \ln(2)}{x(x^2+1)}~.
\nonumber\\ &&
\end{align}
Here the following procedure will be useful.  Let us distinguish the
letters using indices $\alpha$ and denote the corresponding rational
functions with $f_{\alpha}(x)$.  One can form new letters by scaling
the argument of the rational functions with a variable $y$, cf.~Eq.~(\ref{eq:scaleLetter}).  
If one such letter is built into a cyclotomic
HPL with argument $x=1$, there is an algorithm for removing the
parameter $y$ from the index, such that it occurs in the argument. 

At first, by virtue of the shuffle algebra, the weighted letter is
brought to the right-most position.  Then indexing general rational
letters with $\alpha_i$, $i=1,...,n$, we find the algorithm
\begin{eqnarray}
 H_{\alpha_1,...,\alpha_{n-1},[\alpha_n,y]} (1)
 &=&
 \int_0^1 dx_1\;f_{\alpha_1}(x_1)
 ...
 \int_0^{x_{n-2}} dx_{n-1}\;
 f_{\alpha_{n-1}} (x_{n-1})
 \int_0^{x_{n-1}} dx_n\;
 f_{\alpha_n}(yx_n)
 \N\\
 &=&
 \frac{1}{y}
 \int_0^{y} dx_n\;
 \int_0^1 dx_1 f_{\alpha_1}(x_1)
 ...
\N\\&&
 ...
 \int_0^{x_{n-2}} dx_{n-1}
 x_{n-1}
 f_{\alpha_{n-1}}(x_{n-1})
 f_{\alpha_n}(x_{n-1}x_n)
\N\\
 &=&
 \text{``cycl. HPLs''}
 +
 \frac{1}{y}
 \int_0^{y} dx_n\;
 f_{\beta_n}(x_n)
 H_{\alpha_1,...,\alpha_{n-2},[\tilde{\alpha}_n,x_n]}(1)
\comma
\end{eqnarray}
where a partial fraction decomposition is performed in the last step.  After that
the formula may be recursively applied where the final step is
obviously
\begin{eqnarray}
 H_{[\alpha_n,x_2]}(1) = \frac{1}{x_2} H_{\alpha_n}(x_2)
\period
\end{eqnarray}
So the result is a multivariate polynomial in iterated integrals
of arguments $1,y$. This procedure produces also letters $1/(1-x)$,
which introduce branch points at $y=1$. However, considering the
integration contour of the iterated integrals infinitesimally away
from the real axis does not affect the algorithm introduced above. In
this sense, the iterated integrals may be analytically continued, as
described in \cite{Remiddi:1999ew} and implemented in \HSums{} \cite{Harmonicsums,Ablinger:2011te,Ablinger:2013cf}.
Thus they can always be expressed as iterated integrals with
arguments in $[0,1]$.
\subsection{The Final Integral}

\vspace*{1mm}
\noindent
Once the sums are performed, i.e.\ written in terms of iterated
integrals, the remaining task is to perform the last integration,
which carries the nontrivial dependence on $N$.  However, the integral
does not just represent a Mellin-transform, but it contains rational
functions $R(T)\in \{1/(1+T^2),T^2/(1+T^2)\}$, which are raised to the
power $N$. It therefore seems most natural to consider the generating
function of the sequence in $N$, and to introduce the
corresponding tracing parameter $\kappa$ in the following way
\begin{align}
 \sum_{N=0}^{\infty}
 (\kappa R(T))^N
 =
 \frac{1}{1-\kappa R(T)}
\period
\end{align}

The integral over $T$ from $0$ to $1$ is performed in two steps\,: First
a primitive is calculated for the integral in terms of iterated
integrals.  Then the limits $T \rightarrow 1$ and $T\rightarrow 0$ are
computed.  This procedure introduces additional letters into the
otherwise cyclotomic alphabet of HPLs, namely
\begin{align}
\label{eq:eqmassAddLett}
 \frac{1}{1+g(\kappa)T^2} = f_{(4,0)}\left(\sqrt{g(\kappa)}T\right),
 \quad
 \frac{T}{1+g(\kappa)T^2} 
 = \frac{1}{\sqrt{g(\kappa)}} f_{(4,1)}\left(\sqrt{g(\kappa)}T\right)
\comma
\end{align}
with $g(\kappa) \in \{(1-\kappa),(1-\kappa)^{-1}\}$.  Obviously this leads
again to re-scaled letters, and one can use the algorithm from above to
transform the emerging cyclotomic HPLs at 1 with weighted letters into
cyclotomic HPLs with unweighted letters and a function of $\kappa$ in
the argument.  It is not hard to see that the functions occurring in
the arguments of these HPLs are the functions $g(\kappa)$ from above. 

The limit $T\rightarrow 0$ has to be taken carefully to cancel factors of
$1/T$.  Therefore
a Taylor expansion is performed.  In many cases, relations similar to
Eq.~(\ref{eq:CS11relH}) are used, reading them from right to left in
order to obtain the Taylor series.  However, such relations are not
implemented in \HSums{} for the additional (weighted) letters of
Eq.~(\ref{eq:eqmassAddLett}).  This is due to the requirement of
special assumptions on the values of $g(\kappa)$.  We rather use an
easy trick to obtain the Taylor series of cyclotomic HPLs extended by
the above letter, using the fact that the above letter can be
factorized over the complex numbers
\begin{align}
  \frac{1}{1+g(\kappa)T^2}
  =
  \frac{1}{2}
  \left(
  \frac{1}{1 + i\sqrt{g(\kappa)}T}
  +
  \frac{1}{1 - i\sqrt{g(\kappa)}T}
  \right)
\period
\end{align}
Then these linear letters are treated like the letter
\begin{align}
  \frac{1}{a+T}
\end{align}
from the alphabet of multiple polylogarithms \cite{Ablinger:2013cf}, treating
$a$ as real and positive.  For the cyclotomic HPLs extended by one
such letter, the Taylor series expansions can be derived
\cite{Ablinger:2013cf,Ablinger:2011te}.  Finally, the imaginary factors $i\sqrt{g(\kappa)}$ are
re-substituted.  The results are checked to be regular at $T=0$ and
thus the limit can be taken.

Once the last Feynman parameter integral is performed, we need to find
the $N^{\rm th}$ coefficient of the Taylor expansion in $\kappa$.  For this we
would like to make use of methods applicable to HPLs and cyclotomic HPLs 
which are implemented in the package \HSums{} \cite{Harmonicsums,Ablinger:2011te,Ablinger:2013cf}.
It is therefore necessary to make sure that the dependence on $\ln(\kappa)$ cancels.  
These terms can be eliminated using argument
transformations and algebraic relations \cite{Blumlein:2003gb} of the (cyclotomic)
HPLs.\footnote{Note that recently methods for the automatic
extractions of logarithmic parts were implemented in \HSums{}.}

At first, the arguments are mapped back into the interval $[0,1]$
\begin{align}
  H_{\vec{\alpha}}\left(\frac{1}{\sqrt{1-\kappa}} \right)
  =
  \sum_{\vec{\beta}}
  a_{\vec{\beta}}
  H_{\vec{\beta}}\left(\sqrt{1-\kappa}\right)
\comma
\end{align}
where the length of the list $\vec{\beta}$ is bounded by the length of
$\vec{\alpha}$, and the $a_{\vec{\beta}}$ are integer coefficients.  
Relations of this kind can be obtained algorithmically and are
implemented for all cyclotomic HPLs in the package \HSums{}.

Then the square roots are removed from the arguments, as far as
possible.  For this step one makes use of the fact that all cyclotomic
HPLs with arguments $x^2$ can be rewritten in terms of cyclotomic HPLs with 
arguments $x$.  These transformations can be 
inverted, so that
(cyclotomic) HPLs which contain the letter
$f_{(1,0)}(x)=\frac{1}{x-1}$ and the argument $\sqrt{1-\kappa}$ are
mapped onto (cyclotomic) HPLs with argument $1-\kappa$ and
(cyclotomic) HPLs without the letter $f_{(1,0)}$, i.e.\
\begin{align}
  H_{\vec{\alpha}}\left(\sqrt{1-\kappa}\right)
  =
  \sum_{\vec{\beta}}
  b_{\vec{\beta}}
  H_{\vec{\beta}}(1-\kappa)
  +
  \sum_{\vec{\gamma}}
  c_{\vec{\gamma}}
  H_{\vec{\gamma}}\left(\sqrt{1-\kappa}\right)
\comma
\end{align}
where in the vector $\vec{\alpha}$ there is an index $(1,0)$.  The
length of $\vec{\beta}$ is again bounded by the length of
$\vec{\alpha}$, and $\vec{\gamma}$ is free of the index $(1,0)$.

This reduction is, however, not complete  so it is introduced by
constructing a basis of HPLs w.r.t.\ the shuffle relations as well as the
relations of squared arguments.  It is a sign of a proper
Laurent-series that after the reduction to such a basis the remaining
(cyclotomic) HPLs involving the letter $f_{(1,0)}$ and with argument
$\sqrt{1-\kappa}$ will cancel.  

The last step to properly cancel logarithmic parts is to write all
$\ln(\kappa)$ parts explicitly, using the flip relation
\begin{align}
  H_{\alpha}(1-\kappa)
  =
  \sum_{\vec{\eta}}
  d_{\vec{\eta}}
  H_{\vec{\eta}}(\kappa)
\period
\end{align}
In the present case, this relation has to be applied only to HPLs with
letters from the alphabet
\begin{align}
 \left\{f_{0}(x)  = \frac{1}{x},
   f_{1}(x)  = \frac{1}{1-x},
   f_{-1}(x) = \frac{1}{1+x}
 \right\}
\period
\end{align}
This subset is not closed under the flip $x\rightarrow (1-x)$, so the
property
\begin{align}
  f_{-1}(1-x)=\frac{1}{2-x} =: f_{2}(x)
\end{align}
will lead to multiple polylogarithms \cite{Ablinger:2013cf} in the result. 

Nevertheless, the representation is standardized so that indeed all
dependencies on $\ln(\kappa)$ cancel.  The remaining HPLs fit into the
alphabet
\begin{eqnarray}
\small
  \Biggl\{
    f_0(x)=\frac{1}{x},
    f_1(x)=\frac{1}{1-x},
    f_{-1}(x)=\frac{1}{1+x},
    f_2(x)=\frac{1}{2-x},
    f_{(4,0)}(x)=\frac{1}{1+x^2},
\nonumber\\ 
    f_{(4,1)}(x)=\frac{x}{1+x^2}
  \Biggr\}
\comma
\end{eqnarray}
where the letters $f_0, f_{-1}, f_{(4,0)}, f_{(4,1)}$ occur in HPLs
with arguments $\sqrt{1-\kappa}$, and letters $f_1, f_{-1}, f_{(4,0)},
f_{(4,1)}$ lead to HPLs with argument $\kappa$.\footnote{For another algorithm to deal with
polynomial denominators based on the co-product of the associated Hopf-algebra, see Ref.~\cite{vonManteuffel:2013vja}.}

The result thus obtained has a Taylor expansion in $\kappa$ around 0. The remaining step
to obtain the all-$N$ result is to extract the $N^{\rm th}$ coefficient of
the corresponding Taylor series.  This can be done analytically term by
term, using expansions of individual factors and calculating their
Cauchy products, as well as by deriving difference equations which are
solved in terms of indefinite nested sums.  Also these methods are
available through the packages \HSums{} and \CSsigma{}.

As a result of this procedure one obtains a large expression in terms
of sums of higher depth, involving definite and indefinite sums and
products.  To obtain a minimal representation, the package \EvalMS{}
and \CSsigma{} can be applied, in order to represent these objects in
terms of indefinite nested sums, and in order to eliminate all
relations among these indefinite nested sums and products to obtain
a basis-representation.

\subsection{Operator Insertions on External Vertices}
The class of graphs with two massive fermion lines of the same mass
also includes graphs with operator insertions on external gluon
vertices.  In the scalar case these graphs are directly related to
graphs with operator insertions on a line, see \cite{ABHKSW12}
for similar properties used in the calculation of ladder graphs.  

The idea carries over to the physical case, but there are no simple
relations among graphs. Instead if a method is
known for the calculation of certain graphs with operator insertions
on lines, then the same methods apply for the graphs with operator
insertions on external gluon vertices.

The reason lies in the structure of the Feynman rule for the operator
insertion of a gluon vertex, which can be taken from
\cite{BBK09NPB,Klein:2009ig} 
\begin{align}
\label{eq:Vabcmnl}
  V^{abc}_{\mu\nu\lambda}(q_1, q_2, q_3)
  =\;&
  -ig\frac{1+(-1)^N}{2} f^{abc}
  \Biggl[
\N\\&
    t^{3g}_{\mu\nu\lambda}(q_1,q_2,q_3) (\Delta.q_1)^{N-2}
    +
    \tau^{3g}_{\mu\nu\lambda}(q_1,q_2,q_3)
    \sum_{j=0}^{N-3} (-\Delta.q_1)^{j} (\Delta.q_2)^{N-3-j}
\N\\&
    +
    t^{3g}_{\nu\lambda\mu}(q_2,q_3,q_1) (\Delta.q_2)^{N-2}
    +
    \tau^{3g}_{\nu\lambda\mu}(q_2,q_3,q_1)
    \sum_{j=0}^{N-3} (-\Delta.q_2)^{j} (\Delta.q_3)^{N-3-j}
\N\\&
    +
    t^{3g}_{\lambda\mu\nu}(q_3,q_1,q_2) (\Delta.q_3)^{N-2}
    +
    \tau^{3g}_{\lambda\mu\nu}(q_3,q_1,q_2)
    \sum_{j=0}^{N-3} (-\Delta.q_3)^{j} (\Delta.q_1)^{N-3-j}
  \Biggr]
\comma
\end{align}
with
\begin{align}
  t^{3g}_{\mu\nu\lambda}(q_1,q_2,q_3)
  =\;&
  (\Delta_\nu g_{\lambda \mu}-\Delta_\lambda g_{\mu\nu})
  \Delta.p_1 
  +
  \Delta_\mu(p_{1,\nu}\Delta_\lambda-p_{1,\lambda}\Delta_{\nu})
\comma
\N\\
  \tau^{3g}_{\mu\nu\lambda}(q_1,q_2,q_3)
  =\;&
  \Delta_{\lambda}
  \Bigl[
    \Delta.p_1 p_{2,\mu}\Delta_\nu
   +\Delta.p_2 p_{1,\nu}\Delta_\mu
   -\Delta.p_1 \Delta.p_2 g_{\mu\nu}
   -p_1.p_2 \Delta_\mu \Delta_\nu
  \Bigr]
\period
\end{align}
In this notation, the summands in the left column of Eq.\
(\ref{eq:Vabcmnl}) all behave like operator insertions on lines.
Furthermore, if $q_1=p$ is the external momentum then the first and
last summand in the second column behave like insertions on lines too,
but here in addition the result is subject to a finite sum of the form
\begin{align}
 \sum_{j=0}^{N-3} (-\Delta.p)^j (\Delta.p)^{N-3-j} f(N-3-j) 
 =\;&
 (\Delta.p)^{N-3} \sum_{j=0}^{N-3} (-1)^j f(N-3-j) 
\N\\
 =\;&
 (-\Delta.p)^{N-3} \sum_{j=0}^{N-3} (-1)^j f(j) 
\period
\end{align}
The remaining summand (second term, right column) can be summed on the
level of Feynman rules, and using $q_2+q_3=-q_1=-p$ one finds
\begin{align}
  \sum_{j=0}^{N-3} (-\Delta.q_2)^{j} (\Delta.q_3)^{N-3-j}
  =\;
  \frac{1}{\Delta.p}
  \Bigl[
    (-\Delta.q_2)^{N-2}
    -
    (\Delta.q_3)^{N-2}
  \Bigr]
\period
\end{align}
In this way, the operator insertion on an external vertex is related to
operator insertions on internal lines.  However, a direct relation
between a graph with a vertex insertion and the corresponding graphs
with line insertions does not follow from this consideration, due to
the presence of the tensors $t_{\mu\nu\lambda}^{3g}$ and
$\tau_{\mu\nu\lambda}^{3g}$.
\subsection{Integration by parts and differential equations}
\label{sec:3.n}

\begin{center}
\begin{figure}[H]
\centering
\begin{center}
\begin{minipage}[c]{0.35\linewidth}
     \includegraphics[width=1\textwidth]{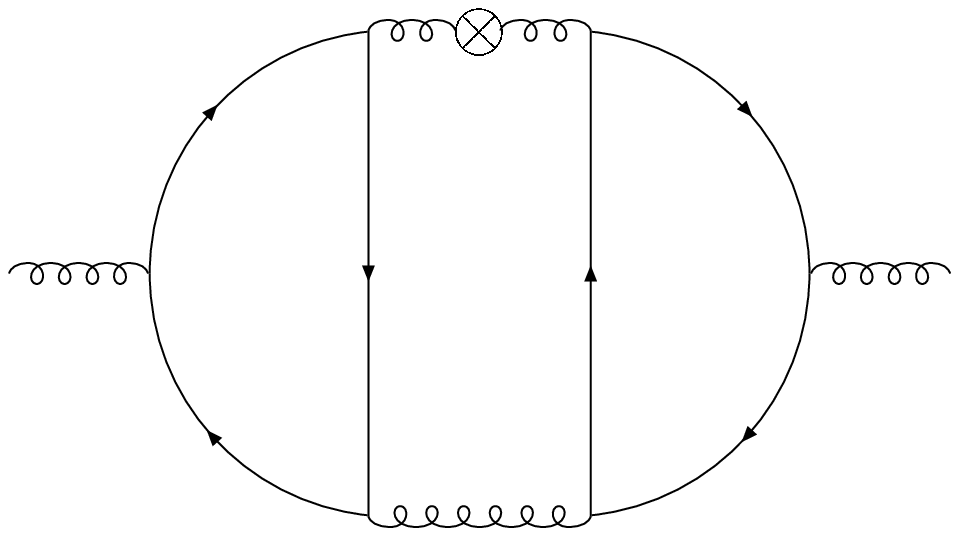}
\vspace*{-4mm}
\begin{center}
{\footnotesize (a)}
\end{center}
\end{minipage}
\hspace*{7mm}
\begin{minipage}[c]{0.33\linewidth}
     \includegraphics[width=1\textwidth]{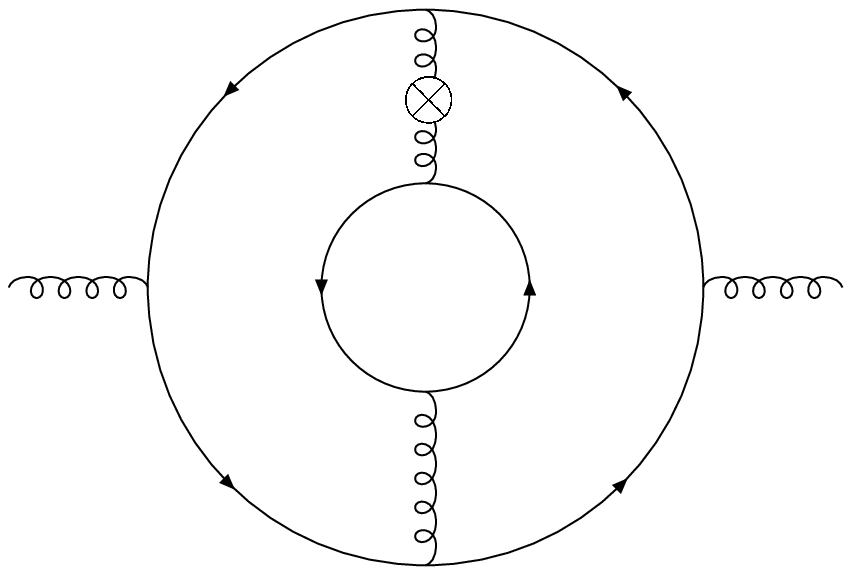}
\vspace*{-7mm}
\begin{center}
{\footnotesize (b)}
\end{center}
\end{minipage}

\vspace*{6mm}
\begin{minipage}[c]{0.33\linewidth}
     \includegraphics[width=1\textwidth]{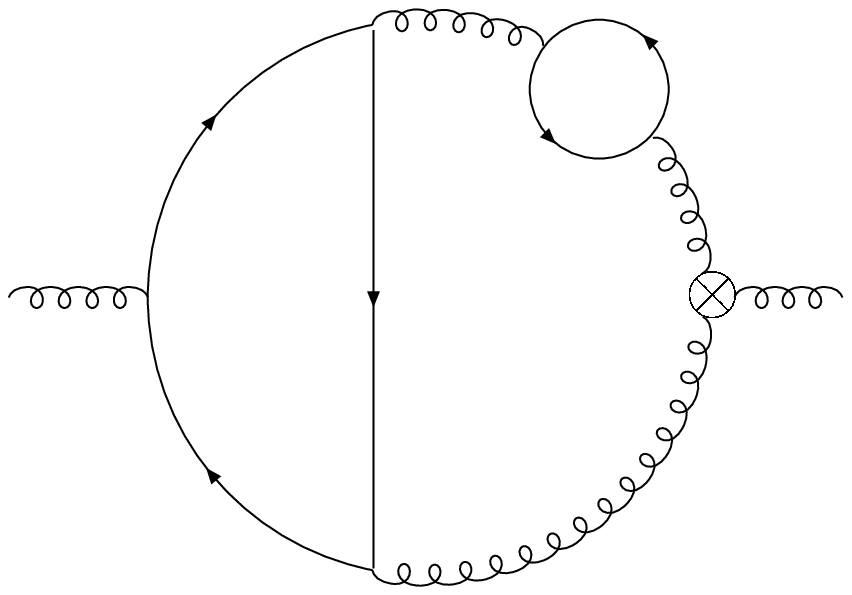}
\vspace*{-7mm}
\begin{center}
{\footnotesize (c)}
\end{center}
\end{minipage}
\hspace*{7mm}
\begin{minipage}[c]{0.33\linewidth}
     \includegraphics[width=1\textwidth]{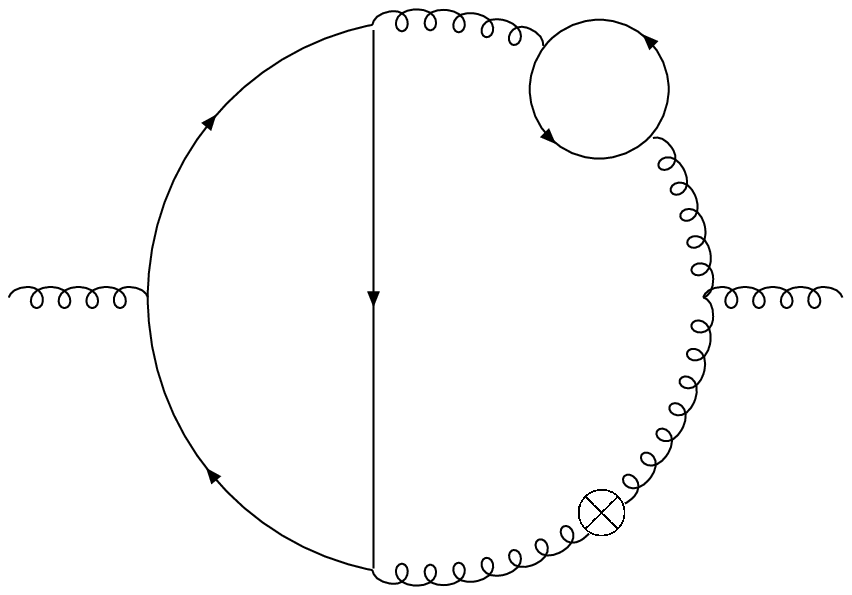}
\vspace*{-7mm}
\begin{center}
{\footnotesize (d)}
\end{center}
\end{minipage}
\caption{\sf Diagrams calculated using differential (difference) equations. Here the masses for both fermion loops are equal.}
\label{diffeqdiags}
\end{center}
\end{figure}
\end{center}

\vspace*{1mm}
\noindent
The diagrams shown in Figure~\ref{diffeqdiags} turned out to be too cumbersome to be calculated with the methods described 
before. 
For this reason, these diagrams were computed using a different approach. For each diagram, a {\tt Form} program 
\cite{FORM} was written in order to replace the propagators and vertices from the output of {\tt QGRAF} \cite{Nogueira:1991ex}
by the corresponding Feynman rules. Further it introduces the corresponding projector for the Green's function under 
consideration and 
performs the Dirac-algebra in the numerator. After this, each diagram ends up being expressed as a linear combination of 
scalar integrals, which were then reduced using integration by parts to master integrals using the program {\tt Reduze2}
\cite{vonManteuffel:2012np}\footnote{The package {\tt Reduze2} uses the packages {\tt Fermat} \cite{FERMAT}
and {\tt Ginac} \cite{Bauer:2000cp}.}. This is a {\tt  C++} program based on Laporta's algorithm \cite{Laporta:2001dd}, 
and has been adapted to the case were we have operator insertions in the integrals. 
In total, sixteen master integrals 
were needed in order to calculate these diagrams. Eleven of them have the general form
\begin{equation}
J^D_{\nu_1, \ldots, \nu_9}(N) = \int dk \,
\frac{(\Delta.k_3)^N}{D_1^{\nu_1} D_2^{\nu_2} \cdots D_9^{\nu_9}} \, ,
\label{int0}
\end{equation}
where we use the shorthand notation
\begin{equation}
\int dk \rightarrow \int \frac{d^D k_1}{(2\pi)^D} \frac{d^D k_2}{(2\pi)^D} \frac{d^D k_3}{(2\pi)^D} \, ,
\end{equation}
and
\begin{eqnarray}
& D_1 = k_1^2-m^2, \quad D_2 = (k_1-p)^2-m^2, \quad D_3 = k_2^2-m^2, \quad D_4 = (k_2-p)^2-m^2, & \nonumber \\
& D_5 = k_3^2, \quad D_6 = (k_3-k_1)^2-m^2, \quad D_7 = (k_3-k_2)^2-m^2, & \nonumber \\
& D_8 = (k_1-k_2)^2, \quad D_9 = (k_3-p)^2 \, . &
\end{eqnarray}

The superscript $D$ has been included in $J^D_{\nu_1, \ldots, \nu_9}(N)$ in order to make explicit the dependence on 
the dimension $D$. The eleven integrals of this type are then
\begin{eqnarray}
J^D_1(N) &=& J^D_{0,1,1,0,0,1,1,0,0}(N), \label{J1} \\
J^D_2(N) &=& J^D_{0,2,1,0,0,1,1,0,0}(N), \label{J2} \\
J^D_3(N) &=& J^D_{0,3,1,0,0,1,1,0,0}(N), \label{J3} \\
J^D_4(N) &=& J^D_{1,1,1,0,0,1,1,0,0}(N), \label{J4} \\
J^D_5(N) &=& J^D_{2,1,1,0,0,1,1,0,0}(N), \label{J5} \\
J^D_6(N) &=& J^D_{1,1,0,1,0,1,1,0,0}(N), \label{J6} \\
J^D_7(N) &=& J^D_{2,1,0,1,0,1,1,0,0}(N), \label{J7} \\
J^D_8(N) &=& J^D_{0,1,0,1,1,1,1,0,0}(N), \label{J8} \\
J^D_9(N) &=& J^D_{1,0,1,0,0,1,1,0,1}(N), \label{J9} \\
J^D_{10}(N) &=& J^D_{1,1,0,1,1,1,1,0,0}(N), \label{J10} \\
J^D_{11}(N) &=& J^D_{1,1,1,0,0,1,1,0,1}(N). \label{J11} 
\end{eqnarray}
The other five master integrals are
\begin{eqnarray}
J^D_{12} &=& \int dk \, \frac{(\Delta . k_1)^N}{(k_1-p)^2 (k_3^2-m^2) [(k_3-k_1)^2-m^2] [(k_3-k_2)^2-m^2]}, \\
J^D_{13} &=& \int dk \, \frac{(\Delta . k_1)^N}{k_1^2 [(k_3-k_1)^2-m^2] [(k_3-k_2)^2-m^2] [(k_3-p)^2-m^2]}, \\
J^D_{14} &=& \int dk \, \frac{(\Delta . k_3)^N}{(k_3^2-m^2) [(k_3-k_1)^2-m^2] [(k_3-k_2)^2-m^2] [(k_3-p)^2-m^2]}, \\
J^D_{15} &=& \int dk \, \frac{1}{(k_3^2-m^2) [(k_3-k_1)^2-m^2] [(k_3-k_2)^2-m^2]}, \\
J^D_{16} &=& \int dk \, \frac{1}{(k_1^2-m^2) (k_2^2-m^2) [(k_3-k_1)^2-m^2] [(k_3-k_2)^2-m^2]}~.
\end{eqnarray}
Notice that integrals $J^D_{15}$ and $J^D_{16}$ are just constants w.r.t. $N$. The integrals $J^D_{12}$, $J^D_{13}$ and 
$J^D_{14}$
yield Feynman parameter integrals that can be performed in terms of Beta-functions. We obtain
\begin{eqnarray}
J^D_{12}(N) &=& -i \Gamma(1-D/2) \Gamma(3-D) \frac{\Gamma(2-D/2)^2}{\Gamma(4-D)} \frac{\Gamma(D/2-1) 
\Gamma(N+1)}{\Gamma(N+D/2)} \, ,
\\
J^D_{13}(N) &=& -i \Gamma(1-D/2) \Gamma(3-D) \frac{\Gamma(2-D/2)^2}{\Gamma(4-D)} \frac{1}{N-1+D/2} \, ,
\\
J^D_{14}(N) &=& -i \frac{\Gamma(1-D/2)^2}{N+1} \Gamma\left(2 - D/2\right) \, ,
\end{eqnarray}
where we have set the mass $m$, $\Delta . p$ and spherical factors to 1 for simplicity.

Any given scalar integral will be written as a linear combination of these master integrals. Since the coefficients of 
these linear combinations may contain poles in $\varepsilon = D-4$, the master integrals may need to be expanded to higher 
orders in $\varepsilon$ accordingly, in order to get the corresponding scalar integrals up to order $\varepsilon^0$.

The integrals $J^D_1(N), \ldots, \, J^D_{11}(N)$ were calculated using the differential equations method \cite{DEQ}. This 
method 
has been applied successfully to many problems where Feynman integrals depending on one or more invariants appear. The idea 
is to take derivatives of the master integrals with respect to these invariants and re-express the result in terms of the 
master integrals themselves. This leads to a system of differential equations that can then be solved. In 
the present case, the integrals depend on the invariants $m^2$ and $\Delta . p$. However, the dependence of the 
integrals on these invariants is trivial. They are just proportional to $(\Delta . p)^N$ and $(m^2)^{-\nu+\frac{3}{2} 
D}$. Here $\nu$ is the sum of powers of propagators. Therefore, taking derivatives with respect to these invariants 
does not lead to any new information. The integrals have the form
\[
F(N) (m^2)^{-\nu+\frac{3}{2} D}  (\Delta . p)^N, 
\]
and it is actually the calculation of the function $F(N)$ expanded in $\varepsilon = D-4$
that is non-trivial. One might think about taking derivatives with respect to $N$, but this changes the structure of the 
integrals in a way that does not allow the application of the differential equations method. In view of this, we introduce 
a new parameter $x$ and rewrite the operator insertion in the following way
\begin{equation}
(\Delta . k_3)^N \to \sum_{N=0}^{\infty} x^N (\Delta . k_3)^N = \frac{1}{1-x \Delta . k_3} \, .
\label{xtrick}
\end{equation}
By doing this, we trade the dependence of the integrals on $N$ by a dependence on $x$, and the operator insertion becomes 
a denominator that can be treated as an additional artificial propagator. In fact, it is in this $x$-representation of the 
integrals in which all the reductions to master integrals are performed using {\tt Reduze2}. Laporta's algorithm requires 
integrals to have definite powers of propagators, and although it may be possible to express $\Delta . k_3$ in terms 
of inverse powers of propagators (by taking $\Delta$ as an external momentum), one faces the problem that the power $N$ in 
the operator insertion $(\Delta . k_3)^N$ is arbitrary. By turning the operator insertion into an artificial 
propagator as in Eq.~(\ref{xtrick}), we circumvent this difficulty. 

Let us define
\begin{equation}
\hat{J}^D_i(x) = \sum_{N=0}^{\infty} x^N J^D_i(N) = \int dk \, \frac{1}{D_1^{\nu_1} \cdots D_9^{\nu_9} 
\left(1-x \Delta . k_3\right)} \, .
\label{JN2Jx}
\end{equation}
We can now take derivatives with respect to $x$.\footnote{In the following we will drop the hat in Eq.~(\ref{JN2Jx}) again, as 
it is clear when we refer to a function depending on the parameter $x$ or the Mellin variable $N$ by the respective argument.} 
This will raise the power of the artificial propagator, leading to integrals 
that can then be reduced, and as usual, a system of differential equations is generated. For example, the first three 
integrals in Eqs. (\ref{J1})-(\ref{J3}), namely, $J^D_1$, $J^D_2$ and $J^D_3$ form the following closed system together with 
the constant integrals $J^D_{15}$ and $J^D_{16}$
\begin{eqnarray}
\label{eq:D1}
\frac{d}{dx} J^{4+\varepsilon}_1(x) &=& 
\frac{\varepsilon x+\varepsilon+2}{2 (x-1) x} J^{4+\varepsilon}_1(x)
-\frac{2}{x-1} J^{4+\varepsilon}_2(x)
-\frac{\varepsilon+2}{2 (x-1) x} J^{4+\varepsilon}_{16} \, ,
\label{deq1}
\\ 
\frac{d}{dx} J^{4+\varepsilon}_2(x) &=& 
-\frac{2 \varepsilon^2 x^2+2 \varepsilon^2 x-\varepsilon^2+9 \varepsilon x-2 \varepsilon+2 x^2+2 x}{2 (x-1) x (\varepsilon+x)} J^{4+\varepsilon}_2(x)
\nonumber \\ &&
+\frac{(\varepsilon+1)^2 (3 \varepsilon+4)}{4 (x-1) (\varepsilon+x)} J^{4+\varepsilon}_1(x)
+\frac{4 \varepsilon}{\varepsilon+x} J^{4+\varepsilon}_3(x)
+\frac{(\varepsilon+2)^3}{8 (x-1) (\varepsilon+x)} J^{4+\varepsilon}_{15}
\nonumber \\ &&
-\frac{(3 \varepsilon+4) (\varepsilon+2) (\varepsilon x+\varepsilon+2 x)}{16 (x-1) x (\varepsilon+x)} J^{4+\varepsilon}_{16} \, ,
\label{deq2}
\\
\label{eq:D3}
\frac{d}{dx} J^{4+\varepsilon}_3(x) &=& 
\frac{2 \varepsilon^2 x^2+\varepsilon^2 x-2 \varepsilon^2-\varepsilon x^2+\varepsilon x-3 x^2+2 x}{2 (x-1) x (\varepsilon+x)} J^{4+\varepsilon}_3(x)
\nonumber \\ &&
+\frac{(\varepsilon+1)^2 (3 \varepsilon+4) (\varepsilon x+\varepsilon-x+1)}{16 (x-1) x (\varepsilon+x)} J^{4+\varepsilon}_1(x)
\nonumber \\ &&
-\frac{\varepsilon^3 \left(2 x^2+5 x\right)-\varepsilon^2 \left(x^2-5 x-9\right)-\varepsilon \left(x^2+5 x-12\right)-4 x+4}{8 (x-1) x (\varepsilon+x)} J^{4+\varepsilon}_2(x)
\nonumber \\ &&
+\frac{(\varepsilon+2)^3 \left(2 \varepsilon^2 x+3 \varepsilon^2-\varepsilon x-2 x\right)}{64 \varepsilon (x-1) x (\varepsilon+x)} J^{4+\varepsilon}_{15}
\nonumber \\ &&
-\frac{(3 \varepsilon+4) (\varepsilon+2) \left(2 \varepsilon^3 x+5 \varepsilon^3+3 \varepsilon^2 x+3 \varepsilon^2-3 \varepsilon x-2 x\right)}{128 \varepsilon (x-1) x (\varepsilon+x)} J^{4+\varepsilon}_{16} \, ,
\label{deq3}
\end{eqnarray}
where we have set $m^2$ and $\Delta . p$ to 1 for simplicity. This system can now be solved, provided the constant 
integrals $J^D_{15}$ and $J^D_{16}$ are previously computed, and a few initial conditions are provided. These
initial conditions will be the values of the integrals and some of their derivatives at $x=0$. Since the $N^{\rm th}$
derivative of 
a given integral $J^D_i(x)$ at $x=0$ is equal to $N! J^D_i(N)$, we see that giving these initial conditions is equivalent to 
giving a few initial values for $J^D_i(N)$. 

When we take the derivatives of the remaining integrals in Eqs. (\ref{J4})-(\ref{J11}), the integrals $J^D_1$, $J^D_2$ and 
$J^D_3$ will also appear on the right hand side of the equations. For example
\begin{equation}
\frac{d}{dx} J^D_4(x) = -\frac{1}{x} J^D_4(x)+\frac{1}{x} J^D_2(x)\, .
\label{deq4}
\end{equation}
So, once we solve the system of Eqs. (\ref{deq1})-(\ref{deq3}), we can substitute the result for $J^D_2(x)$ in 
Eq.~(\ref{deq4}) and solve this equation for $J^D_4(x)$. Likewise, $J^D_4(x)$ will appear on the right hand side of the 
differential equations of the next integrals, etc. We can see that we must solve the system of differential equations
starting with the simplest integrals, and gradually incorporate the results to solve the more complicated ones. This is 
all done with the help of the {\tt Mathematica} packages {\tt Sigma} \cite{SIGMA}, {\tt HarmonicSums} 
\cite{Harmonicsums,Ablinger:2011te,Ablinger:2013cf}, {\tt EvaluateMultiSums}, {\tt SumProduction}  \cite{EMSSP},
and {\tt OreSysG} \cite{Gerhold:02}.
These packages construct a system of difference equations from the 
differential
equations, and then solve for $J^D_i(N)$ directly, instead of $J^D_i(x)$.
For example, one may transform 
the system (\ref{eq:D1})-(\ref{eq:D3})  using Eq. (\ref{JN2Jx}) 
into difference equations.
For large enough values of $N > N_0, N \in \mathbb{N}$ one obtains
\begin{eqnarray}
\label{eq:DIF1}
-(\ep+2 N+2) J_1(N)+(-\ep+2 N-2) J_1(N-1)+4 J_2(N-1) &=&0,
\nonumber\\
\\
16 \ep (\ep-N) J_3(N) 
-8 \big(
   2 \ep^2-\ep-2 N+1
\big) J_3(N-2) 
-8
\big(
   \ep^2-2 N \ep+3 \ep+2 N
\big) J_3(N-1) 
&& \nonumber\\
-(3 \ep+4)(\ep+1)^3 J_1(N) 
+2 (\ep-1) \ep (2 \ep+1) J_2(N-2)
+2 (3 \ep+2)^2 J_2(N) &=& 0, \nonumber\\ \\
\label{eq:DIF3}
4
\big(
   \ep^2+N-1
\big) J_2(N-2)
+2
\big(
   2 \ep^2+2 N \ep+7 \ep-2 N+4
\big) J_2(N-1) 
&& \nonumber\\
+2
\big(
   5 \ep^3+5 \ep^2-5 \ep-4
\big) J_2(N-1)+(1-\ep) (\ep+1)^2 (3 \ep+4) J_1(N-1)
&& \nonumber\\
-2 \ep (\ep+2 N+2) J_2(N)
-(3 \ep+4)(\ep+1)^2 J_1(N-1)
-16 \ep J_3(N-2)+16 \ep J_3(N-1)
&=& 0~.
\nonumber\\
\end{eqnarray}
Here we left out the explicit dependence on the dimension $D = 4 + \ep$ in the functions $J_i$.

Let us discuss now the calculation of the initial values required in order
to solve the differential (difference) equations discussed above. 
These are basically the values of the integrals for a few fixed values of the
Mellin variable $N$. In some cases, these values are needed only up to order $\varepsilon^0$, which can therefore
be obtained using {\tt MATAD} \cite{Steinhauser:2000ry}. More often, the initial values are needed up to higher orders in 
$\varepsilon$, and a different method to obtain them must be used. 
In the following, we describe the
method we used in such cases based on the $\alpha$-parameterization of the integrals. In the present calculation, five
initial values starting from $N=1$ were needed up to order $\varepsilon^2$ for the master integral
\begin{equation}
J^D_1 = J^D_{0,1,1,0,0,1,1,0,0}(N) = \int dk \,
\frac{(\Delta . k_3)^N}{D_2 D_3 D_6 D_7} \, ,
\label{int1}
\end{equation}
and two initial values up to order $\varepsilon$ starting from $N=1$ were needed for 
\begin{equation}
J^D_7 = J^D_{2,1,0,1,0,1,1,0,0}(N) = \int dk \,
\frac{(\Delta . k_3)^N}{D_1^2 D_2 D_4 D_6 D_7} \, .
\label{int2}
\end{equation}

In what follows, the masses appearing in some of the propagators do not play any role, so we will omit them for the time being.
Let us consider the general integral in Eq. (\ref{int0}). Removing the operator insertion (i.e., taking $N=0$) the $\alpha$ 
representation of this integral is given by

\begin{eqnarray}
J^D_{\nu_1, \ldots, \nu_9}(0) &=& \int dk
\prod_{l} \frac{(-1)^{\nu_l}}{\Gamma(\nu_l)} \int_0^{\infty} d\alpha_l \,\, \alpha_l^{\nu_l-1}
\exp\left(\sum_{i} \alpha_i D_i\right) \nonumber \\
&=& \int dk
\prod_{l} \frac{(-1)^{\nu_l}}{\Gamma(\nu_l)} \int_0^{\infty} d\alpha_l \,\, \alpha_l^{\nu_l-1}
\exp\left(\sum_{i,j} A_{i,j} k_i . k_j + 2 \sum_i q_i . k_i\right) \nonumber \\
&\propto&  \prod_{l} \frac{(-1)^{\nu_l}}{\Gamma(\nu_l)} \int_0^{\infty} d\alpha_l \,\, \alpha_l^{\nu_l-1}
\det\left(A\right)^{-D/2} \exp\left(\sum_{i,j} A_{i,j}^{-1} q_i . q_j\right) \, ,
\label{mastereqN0}
\end{eqnarray}
with
\begin{equation}
A = 
\left( {\begin{array}{ccc}
\beta_1+\beta_2+\beta_6+\beta_8 &           -\beta_8              &           -\beta_6              \\
        -\beta_8                & \beta_3+\beta_4+\beta_7+\beta_8 &           -\beta7               \\
        -\beta_6                &           -\beta_7              & \beta_5+\beta_6+\beta_7+\beta_9 \\
  \end{array} } \right) \, ,
\end{equation}
and
\begin{equation}
q_1 = -\beta_2 p, \quad q_2 = -\beta_4 p \quad {\rm and} \quad q_3 = -\beta_9 p \, ,
\end{equation}
where the $\beta_i$'s are defined using the $\theta$-function as
\begin{equation}
\beta_i = \theta\left(\nu_i-\frac{1}{2}\right) \alpha_i \, .
\end{equation}
The product of integrals in the $\alpha$ parameters, and the sum in the exponential in the first and second
lines of Eq. (\ref{mastereqN0}), run over the values of $l$, $i$ and $j$ corresponding to the propagators that are
actually present in the integral under consideration.

We can now introduce the operator insertion in our integrals in the following way, cf. also \cite{Smirnov2006},
\begin{eqnarray}
J^D_{\nu_1, \ldots, \nu_9}(N) &=& \left.\left(\frac{1}{2} \frac{\partial}{\partial r}\right)^N \int dk
\prod_{l} \frac{(-1)^{\nu_l}}{\Gamma(\nu_l)} \int_0^{\infty} d\alpha_l \,\, \alpha_l^{\nu_l-1}
\exp\left(\sum_{i} \alpha_i D_i+2 r \Delta . k_3\right)\right|_{r=0} , \nonumber \\
&\propto&  \left.\left(\frac{1}{2} \frac{\partial}{\partial r}\right)^N 
\prod_{l} \frac{(-1)^{\nu_l}}{\Gamma(\nu_l)} \int_0^{\infty} d\alpha_l \,\, \alpha_l^{\nu_l-1}
\det\left(A\right)^{-D/2} \exp\left(\sum_{i,j} A_{i,j}^{-1} q'_i . q'_j\right)\right|_{r=0} , \nonumber \\
\label{mastereq}
\end{eqnarray}
and now
\begin{equation}
q'_1 = -\beta_2 p, \quad q'_2 = -\beta_4 p \quad {\rm and} \quad q'_3 = -\beta_9 p+r \Delta \, .
\end{equation}

In the case of integral $J^D_1(N)$, we get
\begin{equation}
   A=
  \left( {\begin{array}{ccc}
   \alpha_2+\alpha_6 &         0         &    -\alpha_6      \\
           0         & \alpha_3+\alpha_7 &    -\alpha_7      \\
       -\alpha_6     &      -\alpha_7    & \alpha_6+\alpha_7 \\
  \end{array} } \right) \, ,
\end{equation}
and
\begin{equation}
q'_1 = -\alpha_2 p, \quad q'_2 = 0 \quad {\rm and} \quad q'_3 = r \Delta \, ,
\end{equation}
and one has
\begin{equation}
A^{-1} = \frac{1}{\det(A)} 
\left(
\begin{array}{ccc}
 \alpha_3 \alpha_6+\alpha_7 \alpha_6+\alpha_3 \alpha_7 & \alpha_6
   \alpha_7 & \alpha_3 \alpha_6+\alpha_7 \alpha_6 \\
 \alpha_6 \alpha_7 & \alpha_2 \alpha_6+\alpha_7 \alpha_6+\alpha_2
   \alpha_7 & \alpha_2 \alpha_7+\alpha_6 \alpha_7 \\
 \alpha_3 \alpha_6+\alpha_7 \alpha_6 & \alpha_2 \alpha_7+\alpha_6
   \alpha_7 & (\alpha_2+\alpha_6) (\alpha_3+\alpha_7)
\end{array}
\right).
\end{equation}
If we apply Eq. (\ref{mastereq}) in this case, we obtain
\begin{equation}
J^D_1(N) \propto \prod_{l} \frac{(-1)^{\nu_l}}{\Gamma(\nu_l)} \int_0^{\infty} d\alpha_l \,\, \alpha_l^{\nu_l-1} 
(\alpha_2 \alpha_6)^N (\alpha_3+\alpha_7)^N
\det\left(A\right)^{-(D+2N)/2} \exp\left(\sum_{i,j} A_{i,j}^{-1} q_i . q_j\right) \, ,
\end{equation}
which leads to
\begin{equation}
J^D_1(N) = (\mathbf{2}^+ \mathbf{6}^+)^N (\mathbf{3}^+ + \mathbf{7}^+)^N J^{D+2N}_1(0).
\end{equation}
Here the operator $\mathbf{i}^{+}$ shifts the power of the $i^{\rm th}$ propagator by one, and also multiplies the integral 
by $-\nu_i$, i.e.
\begin{equation}
\mathbf{i}^+ J^D_{\nu_1, \ldots, \nu_i, \ldots, \nu_9} = -\nu_i J^D_{\nu_1, \ldots, \nu_i+1, \ldots, \nu_9} \, .
\end{equation}
The fixed moments for this integral can then be written in terms of scalar integrals with no operator insertion and
shifted values of the dimension and powers of propagators. For example, for $N=1$, $N=2$ and $N=3$ we get
\begin{eqnarray}
J^D_1(1) &=& -J^{D+2}_{0,2,2,0,0,2,1,0,0}(0)-J^{D+2}_{0,2,1,0,0,2,2,0,0}(0), 
\label{Ja1} \\
J^D_1(2) &=& 8 \left[J^{D+4}_{0,3,1,0,0,3,3,0,0}(0)+J^{D+4}_{0,3,2,0,0,3,2,0,0}(0)+J^{D+4}_{0,3,3,0,0,3,1,0,0}(0)\right], 
\label{Ja2} \\
J^D_1(3) &=& -216 \left[J^{D+6}_{0,4,1,0,0,4,4,0,0}(0)+J^{D+6}_{0,4,2,0,0,4,3,0,0}(0)+J^{D+6}_{0,4,3,0,0,4,2,0,0}(0)\right.
\nonumber \\ && \phantom{216 [}
\left.+J^{D+6}_{0,4,4,0,0,4,1,0,0}(0)\right],
\label{Ja3}
\end{eqnarray}
and similar relations for higher values of $N$.

Similarly, it can be shown that
\begin{equation}
J^D_7(N) = (\mathbf{2}^+ \mathbf{4}^+ \mathbf{6}^+ + \mathbf{2}^+ \mathbf{6}^+ \mathbf{7}^+ + \mathbf{1}^+ \mathbf{4}^+ \mathbf{7}^+ +
            \mathbf{2}^+ \mathbf{4}^+ \mathbf{7}^+ + \mathbf{4}^+ \mathbf{6}^+ \mathbf{7}^+)^N J^{D+2N}_7(0) \, .
\end{equation}
For $N=1$ and $N=2$ we have
\begin{eqnarray}
J^D_7(1) &=& -J^{D+2}_{2,1,0,2,0,2,2,0,0}(0) - J^{D+2}_{2,2,0,1,0,2,2,0,0}(0) - J^{D+2}_{2,2,0,2,0,1,2,0,0}(0) 
\nonumber \\ &&
- J^{D+2}_{2,2,0,2,0,2,1,0,0}(0) - 2 J^{D+2}_{3,1,0,2,0,1,2,0,0}(0), 
\label{Jb1} \\
J^D_7(2) &=& 8 \left[J^{D+4}_{2,1,0,3,0,3,3,0,0}(0) + J^{D+4}_{2,2,0,2,0,3,3,0,0}(0) + J^{D+4}_{2,2,0,3,0,2,3,0,0}(0)\right.
\nonumber \\ && \phantom{8 [}
+ J^{D+4}_{2,2,0,3,0,3,2,0,0}(0)+ J^{D+4}_{2,3,0,1,0,3,3,0,0}(0) + J^{D+4}_{2,3,0,2,0,2,3,0,0}(0) 
\nonumber \\ && \phantom{8 [}
+ J^{D+4}_{2,3,0,2,0,3,2,0,0}(0) + J^{D+4}_{2,3,0,3,0,1,3,0,0}(0) + J^{D+4}_{2,3,0,3,0,2,2,0,0}(0) 
\nonumber \\ && \phantom{8 [}
+ J^{D+4}_{2,3,0,3,0,3,1,0,0}(0) +  2 J^{D+4}_{3,1,0,3,0,2,3,0,0}(0) + J^{D+4}_{3,2,0,2,0,2,3,0,0}(0)
\nonumber \\ && \phantom{8 [}
\left.+ 2 J^{D+4}_{3,2,0,3,0,1,3,0,0}(0) + J^{D+4}_{3,2,0,3,0,2,2,0,0}(0) + 3 J^{D+4}_{4,1,0,3,0,1,3,0,0}(0)\right].
\label{Jb2}
\end{eqnarray}
The integrals on the right hand side of Eqs.\ (\ref{Ja1})-(\ref{Ja3}) and Eqs.\ (\ref{Jb1})-(\ref{Jb2}) can all be reduced in 
terms
of the two constant master integrals $J^D_{15}$ and $J^D_{16}$. For example 
\begin{eqnarray}
J^D_{0,2,2,0,0,2,1,0,0}(0) = J^D_{0,2,1,0,0,2,2,0,0}(0) &=& 
\frac{3 (D-3) (D-2) (3 D-10) (3 D-8)}{512 (D-4)} J^D_{16}
\nonumber \\ &&
-\frac{(D-2)^3 (11 D-38)}{256 (D-4)} J^D_{15} \, .
\end{eqnarray}
From Eq. (\ref{Ja1}) we get
\begin{equation}
J^D_1(1) = \frac{3 (D-1) D (3 D-4) (3 D-2)}{256 (D-2)} J^{D+2}_{16}-\frac{D^3 (11 D-16)}{128 (D-2)} J^{D+2}_{15}.
\end{equation}

The integral $J^D_{15}$ is pretty simple and can be obtained for general values of the dimension $D$
\begin{equation}
J^D_{15} = i \Gamma\left(1-\frac{D}{2}\right)^3 \, .
\end{equation}
One can therefore perform without problems the shifts in $D$ for this integral as required 
from Eqs.\ (\ref{Ja1})-(\ref{Ja3}) and Eqs.\ (\ref{Jb1})-(\ref{Jb2}).

The integral $J^D_{16}$ is more complicated. After Feynman parameterization we obtain
\begin{equation}
J^D_{16} = -i \int_0^1 dx \int_0^1 dy \int_0^1 dz \, \Gamma\left(4-\frac{3}{2} D\right)\frac{[x (1-x) y (1-y)]^{-2+D/2} 
[z (1-z)]^{1-D/2}}{\left[\frac{z}{x (1-x)}+\frac{1-z}{y (1-y)}\right]^{4-\frac{3}{2} D}} \, .
\end{equation}
We can now obtain a Mellin-Barnes representation for this integral by splitting the denominator in the equation above using
\begin{equation}
\frac{1}{(A+B)^{\nu}} = \frac{1}{2 \pi i} \int_{\gamma-i \infty}^{\gamma+i \infty} d\sigma \, \frac{\Gamma(-\sigma) 
\Gamma(\sigma+\nu)}{\Gamma(\nu)} \frac{A^{\sigma}}{B^{\sigma+\nu}}
\end{equation}
which leads to
\begin{eqnarray}
J^D_{16} &=& -\frac{1}{2 \pi} \int_{\gamma-i \infty}^{\gamma+i \infty} d\sigma \, \Gamma(-\sigma) 
\Gamma\left(\sigma+4-\frac{3}{2} D\right) 
\frac{\Gamma(-\sigma-1+D/2)^2 \Gamma(\sigma+3-D)^2}{\Gamma(-2 \sigma-2+D) \Gamma(2 \sigma+6-2 D)}
\nonumber \\ && \phantom{\frac{1}{2 \pi i} \int_{-i \infty}^{+i \infty} d\sigma \,}
\times \frac{\Gamma(\sigma+2-D/2) \Gamma(-\sigma-2+D)}{\Gamma(D/2)}\, .
\label{MBrep}
\end{eqnarray}
In this representation, the integral can be calculated with the help of the {\tt Mathematica} package {\tt MB} 
\cite{Czakon:2005rk}. This package finds a value for $\gamma$ and $\varepsilon=D-4$ such that the integral in 
Eq.~(\ref{MBrep}) is well defined. Then it performs an analytic continuation to $\varepsilon \to 0$ and expands in 
$\varepsilon$. After this, we can close the contour to the right or to the left and take residues. This leads
to sums that can be performed with the package {\tt Sigma}. For the different shifts in $D$, we obtain
\begin{eqnarray}
J^{4+\varepsilon}_{16} &=& \frac{16}{\varepsilon^3}-\frac{92}{3 \varepsilon^2}+\frac{6 \zeta_2+35}{\varepsilon}
-\frac{23 \zeta_2}{2}+2 \zeta_3-\frac{275}{12}
+\varepsilon \left(\frac{105 \zeta_2}{8}+\frac{89 \zeta_3}{6}+\frac{57 \zeta_4}{16}-\frac{189}{16}\right)
\nonumber \\ &&
+\varepsilon^2 \biggl[-64 {\rm Li}_4\left(\frac{1}{2}\right)-\frac{8 \ln^4(2)}{3}+\zeta_2 \left(16 \ln^2(2)
+\frac{3 \zeta_3}{4}-\frac{275}{32}\right)+\frac{783 \zeta_2^2}{32}-\frac{525 \zeta_3}{8}
\nonumber \\ &&
+\frac{3 \zeta_5}{10}+\frac{14917}{192}\biggr] + O(\ep^3)\, ,
\\
J^{6+\varepsilon}_{16} &=& -\frac{8}{3 \varepsilon^3}+\frac{911}{135 \varepsilon^2} -\frac{1}{\varepsilon} \left(\zeta_2+\frac{158771}{16200}\right)
+\frac{911 \zeta_2}{360}-\frac{\zeta_3}{3}+\frac{19406231}{1944000}
-\varepsilon \biggl(\frac{158771 \zeta_2}{43200}
\nonumber \\ &&
+\frac{881 \zeta_3}{1080}+\frac{19 \zeta_4}{32}+\frac{1415455691}{233280000}\biggr)
+\varepsilon^2 \biggl[\frac{256 {\rm Li}_4\left(\frac{1}{2}\right)}{45}+\frac{32 \ln^4(2)}{135}
-\frac{17441 \zeta_2^2}{9600}
\nonumber \\ &&
+\zeta_2 \left(-\frac{64 \ln^2(2)}{45}-\frac{\zeta_3}{8}+\frac{19406231}{5184000}\right)
+\frac{810701 \zeta_3}{129600}-\frac{\zeta_5}{20}-\frac{87955543249}{27993600000}\biggr]  
\nonumber\\ &&
+ O(\ep^3)\, ,
\\
J^{8+\varepsilon}_{16} &=& \frac{29}{270 \varepsilon^3}-\frac{432113}{1360800 \varepsilon^2}+\frac{1}{\varepsilon} \left(\frac{29 \zeta_2}{720}+\frac{400656889}{762048000}\right)
-\frac{432113 \zeta_2}{3628800}+\frac{29 \zeta_3}{2160}
\nonumber \\ &&
-\frac{2399678021033}{3840721920000}
+\varepsilon \left(\frac{400656889 \zeta_2}{2032128000}+\frac{26639 \zeta_3}{10886400}+\frac{551 \zeta_4}{23040}+\frac{390635303718683}{716934758400000}\right)
\nonumber \\ &&
+\varepsilon^2 \biggl[-\frac{2048 {\rm Li}_4\left(\frac{1}{2}\right)}{14175}-\frac{256 \ln^4(2)}{42525}+\zeta_2 \left(\frac{512 \ln^2(2)}{14175}+\frac{29
   \zeta_3}{5760}-\frac{2399678021033}{10241925120000}\right)
\nonumber \\ &&
+\frac{71227 \zeta_2^2}{2150400}-\frac{98969999 \zeta_3}{677376000}+\frac{29
   \zeta_5}{14400}-\frac{2591632410097226753}{10840053547008000000}\biggr] + O(\ep^3) \, ,
\\
J^{10+\varepsilon}_{16} &=& -\frac{8}{4725 \varepsilon^3}+\frac{727007}{130977000 \varepsilon^2}+\frac{1}{\varepsilon} \left(-\frac{\zeta_2}{1575}-\frac{24274289111}{2420454960000}\right)
+\frac{727007 \zeta_2}{349272000}
\nonumber \\ &&
-\frac{\zeta_3}{4725}+\frac{16658646415909}{1278000218880000}
+\varepsilon \biggl(-\frac{24274289111 \zeta_2}{6454546560000}+\frac{53651 \zeta_3}{209563200}-\frac{19 \zeta_4}{50400}
\nonumber \\ &&
-\frac{10820372717621142407}{826610541571584000000}\biggr)
+\varepsilon^2 \biggl[\frac{8192 {\rm Li}_4\left(\frac{1}{2}\right)}{5457375}+\frac{1024 \ln^4(2)}{16372125}
-\frac{1011 \zeta_2^2}{7040000}
\nonumber \\ &&
+\zeta_2 \left(-\frac{2048 \ln^2(2)}{5457375}-\frac{\zeta_3}{12600}+\frac{16658646415909}{3408000583680000}\right)
+\frac{21627059753 \zeta_3}{19363639680000}
\nonumber \\ &&
-\frac{\zeta_5}{31500}+\frac{143655436584318407615807}{15275762808242872320000000}\biggr] + O(\ep^3)\, ,
\end{eqnarray}
where we have omitted an overall factor of $i$, and set $m$, $\Delta . p$ and the spherical factor to 1.
We have now all the ingredients required to obtain the initial values. For $J^D_1(N)$ they are
\begin{eqnarray}
J_1^{4+\varepsilon}(1) &=& \frac{8}{\varepsilon^3}
-\frac{46}{3 \varepsilon^2}
+\frac{3 \zeta_2+\frac{35}{2}}{\varepsilon}
-\frac{23 \zeta_2}{4}
+\zeta_3
-\frac{275}{24}
+\varepsilon \left(\frac{57}{80} \zeta_2^2
+\frac{105}{16} \zeta_2
+\frac{89}{12} \zeta_3
-\frac{189}{32}\right)
\nonumber \\ &&
+\varepsilon^2 \biggl[-32 {\rm Li}_4\left(\frac{1}{2}\right)
-\frac{4}{3} \ln^4(2)
+\zeta_2 \left(8 \ln^2(2)
+\frac{3}{8} \zeta_3
-\frac{275}{64}\right)
+\frac{783}{64} \zeta_2^2
-\frac{525}{16} \zeta_3
\nonumber \\ && 
+\frac{3}{20} \zeta_5
+\frac{14917}{384}\biggr] + O(\ep^3),
\\ 
J_1^{4+\varepsilon}(2) &=& \frac{56}{9 \varepsilon^3}
-\frac{298}{27 \varepsilon^2}
+\frac{1}{\varepsilon} \left(\frac{7}{3} \zeta_2+\frac{1873}{162}\right)
-\frac{149}{36} \zeta_2
-\frac{7}{9} \zeta_3
-\frac{11009}{1944}
\nonumber \\ &&
+\varepsilon \biggl[\frac{16}{3} {\rm Li}_4\left(\frac{1}{2}\right)
+\frac{2}{9} \ln^4(2)
+\left(\frac{1873}{432}
-\frac{4}{3} \ln^2(2)\right) \zeta_2
-\frac{137}{80} \zeta_2^2
+\frac{1013}{108} \zeta_3
-\frac{211991}{23328}\biggr]
\nonumber \\ &&
+\varepsilon^2 \biggl[-\frac{332}{9} {\rm Li}_4\left(\frac{1}{2}\right)
-16 {\rm Li}_5\left(\frac{1}{2}\right)
+\frac{2}{15} \ln^5(2)
-\frac{83}{54} \ln^4(2)
-\frac{40645}{1296} \zeta_3
\nonumber \\ &&
+\zeta_2 \left(-\frac{4}{3} \ln^3(2)
+\frac{83}{9} \ln^2(2)
-\frac{7}{24} \zeta_3
-\frac{11009}{5184}\right)
+\left(\frac{14107}{960}
-\frac{34}{5} \ln(2)\right) \zeta_2^2
\nonumber \\ &&
+\frac{391}{30} \zeta_5
+\frac{10107775}{279936}\biggr] + O(\ep^3),
\\
J_1^{4+\varepsilon}(3) &=& \frac{16}{3 \varepsilon^3}
-\frac{80}{9 \varepsilon^2}
+\frac{1}{\varepsilon} \left(2 \zeta_2+\frac{232}{27}\right)
-\frac{10}{3} \zeta_2
-\frac{5}{3} \zeta_3
-\frac{224}{81}
\nonumber \\ &&
+\varepsilon \biggl[ {\rm Li}_4\left(\frac{1}{2}\right)
+\frac{\ln^4(2)}{3}
+\left(\frac{29}{9}
-2 \ln^2(2)\right) \zeta_2
-\frac{117}{40} \zeta_2^2
+\frac{373}{36} \zeta_3
-\frac{10379}{972}\biggr]
\nonumber \\ &&
+\varepsilon^2 \biggl[-\frac{118}{3} {\rm Li}_4\left(\frac{1}{2}\right)
-24 {\rm Li}_5\left(\frac{1}{2}\right)
+\frac{\ln^5(2)}{5}
-\frac{59}{36} \ln^4(2)
+\left(\frac{637}{40}
-\frac{51}{5} \ln(2)\right) \zeta_2^2
\nonumber \\ &&
+\zeta_2 \left(-2 \ln^3(2)
+\frac{59}{6} \ln^2(2)
-\frac{5}{8} \zeta_3
-\frac{28}{27}\right)
-\frac{13235 \zeta_3}{432}
+\frac{779}{40} \zeta_5
+\frac{25324}{729}\biggr] 
\nonumber\\ &&
+ O(\ep^3),
\\
J_1^{4+\varepsilon}(4) &=& \frac{24}{5 \varepsilon^3}
-\frac{566}{75 \varepsilon^2}
+\frac{1}{\varepsilon} \left(\frac{9}{5} \zeta_2+\frac{1697}{250}\right)
-\frac{283}{100} \zeta_2
-\frac{11}{5} \zeta_3
-\frac{15557}{15000}
+\varepsilon \biggl[\frac{48}{5} {\rm Li}_4\left(\frac{1}{2}\right)
\nonumber \\ &&
+\frac{2}{5} \ln^4(2)
+\left(\frac{5091}{2000}
-\frac{12}{5} \ln^2(2)\right) \zeta_2
-\frac{1461}{400} \zeta_2^2
+\frac{26093}{2400} \zeta_3
-\frac{324544}{28125}\biggr]
\nonumber \\ &&
+\varepsilon^2 \biggl[-\frac{4051}{100} {\rm Li}_4\left(\frac{1}{2}\right)
-\frac{144}{5} {\rm Li}_5\left(\frac{1}{2}\right)
+\frac{6}{25} \ln^5(2)
-\frac{4051}{2400} \ln^4(2)
-\frac{960149}{32000} \zeta_3
\nonumber \\ &&
+\zeta_2 \left(-\frac{12}{5} \ln^3(2)
+\frac{4051}{400} \ln^2(2)
-\frac{33}{40} \zeta_3
-\frac{15557}{40000}\right)
+\left(\frac{132357}{8000}
-\frac{306}{25} \ln(2)\right) \zeta_2^2
\nonumber \\ &&
+\frac{1167}{50} \zeta_5
+\frac{3638953021}{108000000}\biggr] + O(\ep^3),
\\
J_1^{4+\varepsilon}(5) &=& \frac{40}{9 \varepsilon^3}
-\frac{892}{135 \varepsilon^2}
+\frac{1}{\varepsilon} \left(\frac{5}{3} \zeta_2+\frac{22541}{4050}\right)
-\frac{223}{90} \zeta_2
-\frac{23}{9} \zeta_3
+\frac{25879}{243000}
+\varepsilon \biggl[\frac{32}{3} {\rm Li}_4\left(\frac{1}{2}\right)
\nonumber \\ &&
+\frac{4}{9} \ln^4(2)
+\left(\frac{22541}{10800}
-\frac{8}{3} \ln^2(2)\right) \zeta_2
-\frac{331}{80} \zeta_2^2
+\frac{96317}{8640} \zeta_3
-\frac{350972423}{29160000}\biggr]
\nonumber \\ &&
+\varepsilon^2 \biggl[-\frac{14779}{360} {\rm Li}_4\left(\frac{1}{2}\right)
-32 {\rm Li}_5\left(\frac{1}{2}\right)
+\frac{4}{15} \ln^5(2)
-\frac{14779}{8640} \ln^4(2)
+\frac{311}{12} \zeta_5
\nonumber \\ &&
+\zeta_2 \left(-\frac{8}{3} \ln^3(2)
+\frac{14779}{1440} \ln^2(2)
-\frac{23}{24} \zeta_3
+\frac{25879}{648000}\right)
+\left(\frac{80923}{4800}
-\frac{68}{5} \ln(2)\right) \zeta_2^2
\nonumber \\ &&
-\frac{30502069}{1036800} \zeta_3
+\frac{114818388451}{3499200000}\biggr] + O(\ep^3)\, .
\end{eqnarray}

The initial values for $J^D_7(N)$ read
\begin{eqnarray}
J^{4+\varepsilon}_7(1) &=& \frac{16}{9 \varepsilon^2}-\frac{49}{18 \varepsilon}
+\frac{2}{3} \zeta_2-\frac{7}{4} \zeta_3+\frac{77}{24}
+\varepsilon \biggl[ 6 {\rm Li}_4\left(\frac{1}{2}\right)
                 +\frac{\ln^4(2)}{4}
                 -\left(\frac{3}{2} \ln^2(2)+\frac{49}{48}\right) \zeta_2
\nonumber \\ &&
                 -\frac{51}{20} \zeta_2^2+\frac{529}{144} \zeta_3-\frac{995}{288}\biggr] + O(\ep^2) \, ,
\\
J^{4+\varepsilon}_7(2) &=& \frac{11}{9 \varepsilon^2}-\frac{65}{36 \varepsilon}
+\frac{11}{24} \zeta_2-\frac{35}{32} \zeta_3+\frac{233}{108}
+\varepsilon \biggl[ \frac{15}{4} {\rm Li}_4\left(\frac{1}{2}\right)
                 +\frac{5}{32} \ln^4(2)
                 -\frac{51}{32} \zeta_2^2
                 +\frac{43}{18} \zeta_3
\nonumber \\ &&
                 -\left(\frac{15}{16} \ln^2(2)+\frac{65}{96}\right) \zeta_2
                 -\frac{6035}{2592}\biggr] + O(\ep^2) \, .
\end{eqnarray}

The solution of Eqs.~(\ref{eq:DIF1})-(\ref{eq:DIF3}) is now obtained in the following way. We uncouple the recurrence system 
using Z\"uricher's algorithm; here we used the package {\tt OreSys} \cite{Gerhold:02}. More precisely we obtain a linear 
recurrence in $J_1(N)$ with polynomial coefficients in $N$ and $\ep$ of order 5. Then activating the recurrency solver
of {\tt Sigma} \cite{Blumlein:2010zv}  and using the above initial values yields the desired solution
expanded in the dimensional parameter $\ep$
\begin{eqnarray}
J_1(N) &=& \frac{8 (N+5)}{3(N+1)}\frac{1}{\ep^3} +
\biggl[-\frac{2
\big(
   9 N^3+40 N^2+41 N+2
\big)}{3 N (N+1)^2}
+\frac{4 (N-1) S_1}{3 (N+1)}\biggr] \frac{1}{\ep^2}\nonumber\\
&&+\biggl[\frac{47 N^5+219 N^4+351 N^3+205 N^2+6 N-4}
{6 N^2 (N+1)^3}
+
\frac{(N-1) S_1^2}{3 (N+1)}
+\frac{(1-N) S_2}{N+1}\nonumber\\
&&+\frac{
\big(
   -9 N^3-4 N^2+13 N+4
\big) S_1}{3 N (N+1)^2}
+
\frac{(N+5) \zeta_2}{N+1}\biggr]
\frac{1}{\ep}\nonumber\\
&&+\frac{-1436 N^4-609 N^3+2 N^2+4 N-8
-133 N^7-678 N^6-1414 N^5
}{24 N^3 (N +1)^4}
\nonumber\\ &&+
\biggl[
   \frac{47 N^5+75 N^4-39 N^3-95 N^2-12 N+8}{12 N^2 (N+1)^3}
   +\frac{(1-N) S_2}
   {2 (N+1)}
\biggr] S_1\nonumber\\
&&+
\frac{(N-1) S_1^3}{18 (N+1)}
+\frac{
\big(
   9 N^3+4 N^2-13 N-4
\big) S_2}{4 N (N+1)^2}
+\frac{2 (N-1) S_{2,1}}{N+1}-
\frac{11 (N-1) S_3}{9 (N+1)}\nonumber\\
&&+
\biggl[
   \frac{-9 N^3-40 N^2-41 N-2}{4 N (N+1)^2}
   +\frac{(N-1) S_1}{2 (N+1)}
\biggr] \zeta
_2\nonumber\\
&&+
\frac{
\big(
   -9 N^3-4 N^2+13 N+4
\big) S_1^2}{12 N (N+1)^2}+\frac{(19-13 N) \zeta_3}{3 (N+1)}+O(\ep),
\label{eq:J1}
\\
J_2(N) &=& \frac{8 (N+3)}{3(N+1)}\frac{1}{\ep^3}+\biggl[-\frac{4
\big(
   3 N^2+8 N+7
\big)}{3 (N+1)^2}
+\frac{4 N S_1}
{3 (N+1)}\biggr]
\frac{1}{\ep^2}+
\biggl[\frac{2
\big(
   5 N^3+15 N^2+17 N+9
\big)}{3 (N+1)^3}\nonumber\\
&&+\frac{N S_1^2}{3 (N+1)}-\frac{N S_2}{N+1}
+\frac{(N+3) \zeta
_2}{N+1}-
\frac{2
\big(
   3 N^2+5 N+1
\big) S_1}{3 (N+1)^2}\biggr]\frac{1}{\ep}\nonumber\\
&&+
\frac{N^4+12 N^3+30 N^2+26 N+5}{3 (N+1)^4}
+
\biggl[
   \frac{5 N^3+18 N^2+20 N+6}{3 (N+1)^3}
   -\frac{N S_2}
   {2 (N+1)}
\biggr] S_1\nonumber\\
&&
+
\frac{
\big(
   3 N^2+5 N+1
\big) S_2}{2 (N+1)^2}+\frac{2 N S_{2,1}}{N+1}-\frac{11 N S_3}{9 (N+1)}
+\frac{N S_1^3}
{18 (N+1)}\nonumber\\
&&+
\biggl[
   \frac{-3 N^2-8 N-7}{2 (N+1)^2}
   +\frac{N S_1}{2 (N+1)}
\biggr] \zeta_2
+\frac{
\big(
   -3 N^2-5 N-1
\big) S_1^2}
{6 (N+1)^2}+\frac{(3-13 N) \zeta_3}{3 (N+1)}
+O(\ep),
\nonumber\\
\label{eq:J2}
\\
J_3(N) &=&  \frac{2 (2 N+5)}
{3(N+1)}\frac{1}{\ep^2}
+\biggl[\frac{-8 N^2-20 N-15}{3 (N+1)^2}
+\frac{(2 N-1) S_1}{3 (N+1)}\biggr]\frac{1}{\ep}
+\frac{(1-2 N) S_2}{4 (N+1)}
\nonumber\\ &&
+\frac{24 N^3+76 N^2+84 N+35}{6 (N+1)^3}
+
\frac{
\big(
   -8 N^2-8 N+3
\big) S_1}{6 (N+1)^2}
+\frac{(2 N-1) S_1^2}{12 (N+1)}
\nonumber\\ && 
+\frac{2^{-2 N-1} (-2 N-1) \binom{2 N}{N}}{N+1}\sum_{{i}_1=1}^N \frac{2^{2 {i}_1}}{\binom{2 {i}_1}{{i}
_1} {i}_1^3}
+
\frac{2^{-2 N-1} (2 N+1) \binom{2 N}{N}}{N+1}\sum_{{i}_1=1}^N \frac{2^{2 {i}_1} S_1
\big(
   {i}_1
\big)}{\binom{2 {i}
_1}{{i}_1} {i}_1^2}\nonumber\\
&&+
\frac{(2 N+5) \zeta_2}{4 (N+1)}-7 \frac{2^{-2 N-1} (2 N+1) \binom{2 N}{N} \zeta_3}{N+1}+O(\ep),
\label{eq:J3}
\end{eqnarray}
which holds for values of $N \geq N_0$. Usually for values of $N < N_0$ additional constants appear. 
Eq.~(\ref{eq:J1}) is valid for $N \geq 1$ and Eqs.~(\ref{eq:J2}, \ref{eq:J3}) for $N \geq 0$.
For the analytic
continuation to $N \in \mathbb{C}$ the additional terms are not relevant.

\section{The \boldmath{$O(\alpha_s^3 T_F^2)$} Contributions to \boldmath{$A_{gg,Q}$}}
\label{sec:4}

\vspace*{1mm}
\noindent
The contributions of $O(\alpha_s^3 T_F^2 C_{F,A})$ to the operator matrix element $A_{gg,Q}$ are obtained as 
respective color-projections from Eq.~(\ref{eqAggQ}). We first consider the contribution to the constant part 
$a_{gg,Q}^{(3)}$ of the unrenormalized OME (\ref{unren}). Defining
\begin{eqnarray}
F(N) = \frac{(2+N+N^2)^2}{(N-1) N^2 (N+1)^2 (N+2)} \equiv F,
\end{eqnarray}
it is given by
\begin{eqnarray}
\label{eq:agg}
\lefteqn{
a_{gg,Q;\rm T_F^2}^{(3)}(N) =} \nonumber\\ && \textcolor{blue}{C_F T_F^2}
\Biggl\{
 \frac{16}{27} F S_1^3
+\frac{16 P_4}{27 (N-1) N^3 (N+1)^3 (N+2)} S_1^2
+\Biggl[-\frac{16}{3} F S_2
\nonumber\\ &&
-\frac{32 P_{10}}{81 (N-1) N^4 (N+1)^4 (N+2) (2 N-3) (2 N-1)} \Biggr] S_1
-\frac{16 P_4}{9 (N-1) N^3 (N+1)^3 (N+2)} S_2
\nonumber\\ &&
-\frac{2 P_{13}}{243 (N-1) N^5 (N+1)^5 (N+2) (2 N-3) (2 N-1)}
- F \left[ \frac{352}{27}   S_3
- \frac{64}{3}  S_{2,1} \right]
\nonumber\\ &&
+\Biggl[\frac{16}{3} F S_1
-\frac{8 P_8}{9 (N-1) N^3 (N+1)^3 (N+2)}\Biggr] \zeta_2
+\frac{P_3}{9 (N-1) N^2 (N+1)^2 (N+2)} \zeta_3
\nonumber\\ &&
-\binom{2N}{N} 
\frac{16 P_5}{3(N-1) N (N+1)^2 (N+2) (2 N-3) (2 N-1)} \frac{1}{4^N}\left(\sum_{i=1}^N \frac{4^i S_1(i-1)}{i^2 \binom{2i}{i}} - 
7 
\zeta_3\right) 
\Biggr\}
\nonumber\\ &&
+\textcolor{blue}{C_A T_F^2} 
\Biggl\{
-\frac{4 P_2}{135 (N-1) N^2 (N+1)^2 (N+2)} S_1^2
+\frac{16 \big(4 N^3+4 N^2-7 N+1\big)}{15 (N-1) N (N+1)} [S_{2,1} - S_3]
\nonumber\\ &&
+\frac{P_{12}}{3645(N-1) N^4 (N+1)^4 (N+2) (2 N-3) (2 N-1)} 
\nonumber\\ &&
-\frac{8 P_{11}}{3645 (N-1) N^3 (N+1)^3 (N+2) (2 N-3) (2 N-1)} S_1
+\frac{4 P_7}{135 (N-1) N^2 (N+1)^2 (N+2)} S_2
\nonumber\\ &&
-\binom{2N}{N} 
\frac{4 P_9}{45 (N-1) N (N+1)^2 (N+2) (2 N-3) (2 N-1)} \frac{1}{4^N} \left(\sum_{i=1}^N 
\frac{4^i S_1(i-1)}{i^2 \binom{2i}{i}} -7 \zeta_3\right)
\nonumber\\ &&
+\Biggl[\frac{4 P_6}{27 (N-1) N^2 (N+1)^2 (N+2)}
-\frac{560}{27} S_1
\Biggr] \zeta_2
+\Biggl[-\frac{7 P_1}{270 (N-1) N (N+1) (N+2)}
\nonumber\\ &&
-\frac{1120}{27} S_1 \Biggr] \zeta_3 \Biggr\},
\end{eqnarray}
with the polynomials $P_i$
\begin{eqnarray}
P_1 &=& 1287 N^4+3726 N^3-3047 N^2-7214 N-2624, 
\\
P_2 &=& 70 N^5+95 N^4-223 N^3-751 N^2-629 N-142,
\\
P_3 &=& -63 N^6-189 N^5-431 N^4-547 N^3-1714 N^2-1472 N-1472,
\\
P_4 &=& 4 N^6+3 N^5-50 N^4-129 N^3-100 N^2-56 N-24,
\\
P_5 &=& 9 N^6+9 N^5-53 N^4+47 N^3+44 N^2-104 N-80,
\\
P_6 &=& 99 N^6+297 N^5+631 N^4+767 N^3+1118 N^2+784 N+168,
\\
P_7 &=& 220 N^6+550 N^5-135 N^4-883 N^3-1621 N^2-1329 N-462,
\\
P_8 &=& 33 N^8+132 N^7+106 N^6-108 N^5-74 N^4+282 N^3+245 N^2+148 N+84,
\\
P_9 &=& 100 N^8+539 N^7+283 N^6-2094 N^5+452 N^4+219 N^3-1495 N^2
\nonumber\\ &&
+712 N+996,
\\
P_{10} &=& 23 N^{10}+136 N^9-221 N^8+388 N^7+1470 N^6+2206 N^5+2192 N^4+2564 N^3
\nonumber\\ &&
+2082 N^2+1008 N+216,
\\
P_{11} &=& 96020 N^{10}+180403 N^9-293651 N^8-563492 N^7+196513 N^6+478087 N^5-194200 N^4
\nonumber\\ &&
-207066 N^3-7470 N^2-38880 N-12960,
\\
P_{12} &=& 149796 N^{12}+481788 N^{11}+4037555 N^{10}+6431215 N^9-710852 N^8-14957774 N^7
\nonumber\\ &&
-21164117 N^6-11167685 N^5+2360450 N^4
+2452488 N^3-1225440 N^2
\nonumber\\ &&
-518400 N+181440,
\\
P_{13} &=& 8868 N^{14}+35472 N^{13}-9409 N^{12}-152862 N^{11}+61883 N^{10}+593774 N^9-379547 N^8
\nonumber\\ &&
-1672874 N^7-807075 N^6
+89818 N^5-325576 N^4-407328 N^3-167688 N^2
\nonumber\\ &&
-21600 N+18144~.
\end{eqnarray} 
Here we use the short-hand notation $S_{\vec{a}}(N) \equiv S_{\vec{a}}$  for the harmonic sums \cite{HSUM}.
The polynomial denominators in Eq.~(\ref{eq:agg}) show evanescent poles at $N = 1/2, 3/2$. However, the function is
continuous at these points, as the expansion around these values shows. This also confirms that the rightmost
pole is located at $N = 1$ as expected for a gluonic quantity in QCD. This also applies to the OME, Eq.~(\ref{eq:OMEOMS}).

In Eq.~(\ref{eq:agg}) the new sum
\begin{eqnarray} 
\label{eq:new1}
T = \frac{1}{4^N} \binom{2N}{N} \left(\sum_{i=1}^N \frac{4^i S_1(i-1)}{i^2 \binom{2i}{i}} - 7
\zeta_3\right),
\end{eqnarray} 
occurs. While all other quantities emerging are known to obey regular asymptotic expansions, it has to be
investigated whether this is also the case for the term (\ref{eq:new1}). Using {\tt HarmonicSums} we obtain
\begin{eqnarray} 
\label{eq:new2}
T &\propto& 
-\frac{4}{N} 
+\frac{7}{9 N^2}
-\frac{79}{450 N^3}
+\frac{937}{22050 N^4}
+\frac{853}{132300 N^5}
-\frac{61807}{3201660 N^6}
-\frac{887287}{2705402700 N^7}
\nonumber\\ &&
+\frac{2650559}{128828700 N^8}
-\frac{419100421}{223388965800 N^9}
-\frac{845167596619}{22580156663064 N^{10}}
+ \Biggl[
-\frac{2}{N}
+\frac{2}{3 N^2}
-\frac{2}{15 N^3}
\nonumber\\ &&
-\frac{2}{105 N^4}
+\frac{2}{105 N^5}
+\frac{2}{231 N^6}
-\frac{54}{5005 N^7}
-\frac{6}{715 N^8}
+\frac{466}{36465 N^9}
+\frac{13646}{969969 N^{10}}
\Biggr] \ln(\bar{N}) 
\nonumber\\ &&
+ O\left(\frac{1}{N^{11}} \ln(\bar{N})\right),
\end{eqnarray} 
with $\bar{N} = N \exp{\gamma_E}$ and $\gamma_E$ denotes the Euler-Mascheroni constant.
A regular asymptotic representation is obtained for Eq.~(\ref{eq:new2}), which is even free of $1/\sqrt{N}$ terms
due to the balanced occurrence of the binomials $\binom{2j}{j}$. Since all other terms
of $O(\alpha_s^3 T_F^2 C_{F,A})$ of $A_{gg,Q}$ contain harmonic sums and rational factors only \cite{LOGS}
the OME behaves the same way, cf.~\cite{ANCONT1}.

It is an interesting question, as to whether new structures, like those in Eq.~(\ref{eq:new1}) compared to the usual 
harmonic sums, can be recognized in studying the minimal difference equations\footnote{Difference equations of this
kind can be generated using the packages {\tt Guess} \cite{GUESS}.}
they obey. For this purpose
we consider the equation for the harmonic sum $S_{2,1}(N)$ at one side and Eq.~(\ref{eq:new1}) on the other side.
The former obeys the difference equation
\begin{eqnarray}
&&-(N+1)^2 (N+2) f_{N}
+(N+2) \big(3 N^2+11 N+11\big) f_{N+1}
\nonumber\\ &&
+\big(-3 N^3-22 N^2-55 N-47\big) f_{N+2}
+(N+3)^3 f_{N+3}
= 0~,
\end{eqnarray}
with the initial values 
\begin{eqnarray}
\big\{f_1 = 1, f_2 = \tfrac{11}{8},f_3 = \tfrac{341}{216},f_4 = \tfrac{2953}{1728}\big\}.
\end{eqnarray}
The term $T$ without the $\zeta_3$-contribution obeys
\begin{eqnarray}
&&
-(2 N+1) (N+1)^2 f_{N}
+(3 N+4) \big(2 N^2+6 N+5\big) f_{N+1}
\nonumber\\ &&
-(N+2) \big(6 N^2+25 N+27\big) f_{N+2}
+2 (N+2) (N+3)^2 f_{N+3}
= 0~,
\end{eqnarray}
with the initial values 
\begin{eqnarray}
\big\{f_1 = 0,f_2 = \tfrac{1}{4}, f_3 = \tfrac{3}{8},f_4 = \tfrac{85}{192}\big\}~.
\end{eqnarray}
Both difference equations are of degree and order three and are of quite similar structure. The different type
of the solutions are therefore hardly recognized ab initio. 

The Mellin inversion of the binomial terms yield \cite{ABRS14}
\begin{eqnarray}
\label{eq:incBinIntegral}
   \sum_{j=1}^N 
   \frac{4^{j} S_{1}(j-1)}
   {\binom{2j}{j} j^2}
  &=&
  \int_0^1 dx\;
  \frac{x^N-1}{x-1}
  \int_x^1 dy\;
  \frac{1}{y \sqrt{1-y}}
  \Bigl[\ln(1-y) - \ln(y) + 2\ln(2) \Bigr],
\\
\frac{1}{4^N} \binom{2N}{N} 
&=& \frac{1}{\pi} \Mvec \left[ \frac{1}{\sqrt{x(1-x)}}\right],
\end{eqnarray}
with the Mellin transform
\begin{eqnarray}
\Mvec
[f(x)](N) = \int_0^1 dx~x^N~f(x)~.
\end{eqnarray}
Therefore the two new letters  \cite{ABRS14}
\begin{eqnarray}
f_{\sf w_1}(x) = \frac{1}{\sqrt{x(1-x)}},~~~~~~~~~f_{\sf w_3}(x) = \frac{1}{x \sqrt{1-x}} 
\end{eqnarray}
appear in the $x$-space representation beyond those forming the usual harmonic polylogarithms \cite{Remiddi:1999ew}.
\subsection{The Operator Matrix Element}

\vspace{1mm}
\noindent
The $O(T_F^2 C_{F,A})$ contribution to the operator matrix $A_{gg,Q}^{(3)}$ is given by
\begin{eqnarray}
\label{eq:OMEOMS}
\lefteqn{A_{gg,Q,\rm T_F^2}^{(3)}(N) =} \nonumber\\ && 
\textcolor{blue}{T_F^2} \Biggl\{ \Biggl\{
\textcolor{blue}{C_F} \frac{80}{9} F 
+ \textcolor{blue}{C_A} \Biggl[\frac{448 \big(N^2+N+1\big)}{27 (N-1) N (N+1) (N+2)}
-\frac{224}{27} S_1\Biggr]
\Biggr\}
\ln^3\left(\frac{m^2}{\mu^2}\right)
\nonumber \\ &&
+\Biggl\{
\textcolor{blue}{C_F} 
\Biggl[\frac{32}{3} F  S_1 
+\frac{8 P_{20}}{9 (N-1) N^3 (N+1)^3 (N+2)}\Biggr]
\nonumber\\ &&
+\textcolor{blue}{C_A}
\Biggl[
\frac{8 P_{19}}{27 (N-1) N^2 (N+1)^2 (N+2)}
-\frac{640}{27} S_1
\Biggr]
\Biggr\} 
\ln^2\left(\frac{m^2}{\mu^2}\right)
\nonumber\\ &&
+\Biggl\{
\textcolor{blue}{C_F} 
\Biggl[\frac{16}{3} [S_1^2 - 3 S_2] F 
-\frac{8 P_{23}}{27 (N-1) N^4 (N+1)^4 (N+2)}
+\frac{32 P_4}{9 (N-1) N^3 (N+1)^3 (N+2)} S_1
\Biggr]
\nonumber\\ &&
+ \textcolor{blue}{C_A} 
\Biggl[
-\frac{2 P_{21}}{27 (N-1) N^3 (N+1)^3 (N+2)}
-\frac{8 P_{18}}{9 (N-1) N^2 (N+1)^2 (N+2)} S_1 \Biggr]
\Biggr\}
\ln\left(\frac{m^2}{\mu^2}\right)
\nonumber\\ &&
-
\textcolor{blue}{C_F} \frac{1}{4^N} \binom{2N}{N} \frac{16 P_5}{3(N-1) N (N+1)^2 (N+2) (2 N-3) (2 N-1)} 
\Biggl[\sum_{j=1}^N \frac{4^j S_1\big(j-1\big)}{\binom{2j}{j} j^2} - 7 \zeta_3 
\Biggr] 
\nonumber\\ &&
- \textcolor{blue}{C_A} \frac{1}{4^N} \binom{2N}{N}
\frac{4 P_{22}}{45 (N-1) N (N+1)^2 (N+2) (2 N-3) (2 N-1)} 
\Biggl[\sum_{j=1}^N \frac{4^k S_1\big(j-1\big)}{\binom{2j}{j} j^2} - 7 \zeta_3
\Biggr] 
\nonumber\\ &&
+\frac{1}{243} \textcolor{blue}{C_F} 
\Biggl[
144 F S_1^3
+\frac{144 P_4}{(N-1) N^3 (N+1)^3 (N+2)} S_1^2
\nonumber\\ &&
+\Biggl[- 1296 F S_2
-\frac{96 P_{10}}{(N-1) N^4 (N+1)^4 (N+2) (2 N-3) (2 N-1)} 
\Biggr] S_1
\nonumber\\ &&
- \frac{189 P_{16}}{(N-1) N^2 (N+1)^2 (N+2)} \zeta_3 
+\frac{8 P_{26}}{(N-1) N^5 (N+1)^5 (N+2) (2 N-3) (2 N-1)} 
\nonumber\\ &&
-\frac{432 P_4}{(N-1) N^3 (N+1)^3 (N+2)} S_2
-3168 F S_3 
+ 5184  F S_{2,1}
-10368 
\zeta_2  \Biggr]
\nonumber\\ &&
+ \textcolor{blue}{C_A} \frac{1}{7290}  \Biggl[
216 \frac{P_{15}}{(N-1)N^2(N+1)^2(N+2)}  S_1^2
\nonumber\\ &&
+7290 \Biggl[\frac{8 P_{24}}{3645 (N-1) N^3 (N+1)^3 (N+2) (2 N-3) (2 N-1)}
-\frac{896}{27} \zeta_3 \Biggr] S_1
\nonumber\\ &&
-189 \frac{P_{14}}{(N-1) N (N+1) (N+2)} \zeta_3
+\frac{2 P_{25}}{(N-1) N^4 (N+1)^4 (N+2) (2 N-3) (2 N-1)}
\nonumber\\ &&
+ 216 \frac{P_{17}}{(N-1)N^2(N+1)^2(N+2)} S_2
-\frac{7776 \big(4 N^3+4 N^2-7 N+1\big)}{(N-1) N (N+1)} [S_3 - S_{2,1}]
\Biggr]
\Biggr\},
\end{eqnarray}
using Eqs.~(\ref{eqAggQ}, \ref{eq:agg}) and the corresponding expressions implied by renormalization from 
Ref.~\cite{LOGS}.
The polynomials $P_i$ read
\begin{eqnarray}
P_{14} &=& 1287 N^4+3726 N^3-2407 N^2-6574 N-1984,
\\
P_{15} &=& 20 N^5+85 N^4+133 N^3+571 N^2+629 N+142,
\\
P_{16} &=& 9 N^6+27 N^5+73 N^4+101 N^3+302 N^2+256 N+256,
\\
P_{17} &=& 40 N^6+100 N^5-135 N^4-433 N^3-1441 N^2-1329 N-462,
\\
P_{18} &=& 40 N^6+114 N^5+19 N^4-132 N^3-147 N^2-70 N-32,
\\
P_{19} &=& 63 N^6+189 N^5+367 N^4+419 N^3+626 N^2+448 N+96,
\\
P_{20} &=& 15 N^8+60 N^7+76 N^6-18 N^5-275 N^4-546 N^3-400 N^2-224 N-96,
\\
P_{21} &=& 27 N^8+108 N^7-1440 N^6-4554 N^5-5931 N^4-3762 N^3-256 N^2
\nonumber\\ &&
+1184 N+480,
\\
P_{22} &=& 100 N^8+539 N^7+283 N^6-2094 N^5+452 N^4+219 N^3-1495 N^2
\nonumber\\ &&
+712 N+996,
\\
P_{23} &=& 219 N^{10}+1095 N^9+1640 N^8-82 N^7-2467 N^6-2947 N^5-3242 N^4-4326 N^3
\nonumber\\ &&
-3466 N^2
-1488 N-360,
\\
P_{24} &=& 22060 N^{10}+29837 N^9-86869 N^8-94588 N^7+64757 N^6+39953 N^5+107890 N^4
\nonumber\\ &&
+78546 N^3
+36630 N^2+38880 N+12960,
\\
P_{25} &=& 145476 N^{12}+468828 N^{11}-697525 N^{10}-2435225 N^9-540932 N^8+3047266 N^7
\nonumber\\ &&
+2170723 N^6-1077965 N^5-2704030 N^4-1889112 N^3-674640 N^2-207360 N
\nonumber\\ &&
-51840,
\\
P_{26} &=& 8340 N^{14}+33360 N^{13}+13051 N^{12}-98742 N^{11}-127865 N^{10}+59578 N^9
\nonumber\\ &&
+195617 N^8
+147746 N^7+91089 N^6+112370 N^5+98404 N^4+59064 N^3
\nonumber\\ &&
+27828 N^2
+7344 N+1296~.
\end{eqnarray}
The analytic continuation of the OME Eq.~(\ref{eq:OMEOMS}) from the even moments $N  = 2n, n \in \mathbb{N}$ 
to the complex plane is obtained using the asymptotic representation for the harmonic sums \cite{ANCONT1,ANCONT}
and Eq.~(\ref{eq:new1}) supplemented by the recursion relations for $N \rightarrow (N-1)$ of Eq.~(\ref{eq:OMEOMS}).

The OME in the $\overline{\rm MS}$ scheme is obtained by the following transformation
\begin{eqnarray}
A_{gg,Q}^{(1),\rm \overline{MS}} -A_{gg,Q}^{(1),\rm OMS} &=&  0,
\\
A_{gg,Q}^{(2),\rm \overline{MS}} -A_{gg,Q}^{(2),\rm OMS} &=& 
\textcolor{blue}{C_F T_F} \frac{8}{3}\left[4-3 \ln\left(\frac{m^2}{\mu^2}\right)\right], 
\\
A_{gg,Q}^{(3),\rm \overline{MS}} -A_{gg,Q}^{(3),\rm OMS} &=& 
\ln^2\left(\frac{m^2}{\mu^2}\right)
\Biggl\{
\textcolor{blue}{C_F} 
\Biggl[
\textcolor{blue}{C_A T_F}
\Biggl[\frac{4 P_{27}}{3 (N-1) N (N+1) (N+2)}
+32 S_1
\Biggr]
\nonumber\\ && 
-\frac{16}{3} (\textcolor{blue}{N_F}+5) \textcolor{blue}{T_F^2}
\Biggr]
-\textcolor{blue}{C_F^2 T_F} 48 F  
\Biggr\}
\nonumber\\ &&
+\ln\left(\frac{m^2}{\mu^2}\right) 
\Biggl\{
\textcolor{blue}{C_F}
\Biggl[
\textcolor{blue}{C_A T_F} 
\Biggl[
\frac{32}{3} S_1
-\frac{4 P_{28}}{9 (N-1) N^2 (N+1)^2 (N+2)}
\Biggr]
\nonumber\\ && 
+\frac{16}{9} (13 \textcolor{blue}{N_F}+29) \textcolor{blue}{T_F^2}
\Biggr]
+\textcolor{blue}{C_F^2 T_F} \frac{4 P_{31}}{(N-1) N^3 (N+1)^3 (N+2)}
\Biggr\}
\nonumber\\ &&
+\textcolor{blue}{C_F} 
\Biggl\{
\textcolor{blue}{C_A T_F} 
\Biggl[
\frac{P_{29}}{9 (N-1) N^2 (N+1)^2 (N+2)}
+ 64 \Biggl[\ln(2) - \frac{1}{3}\Biggr] \zeta_2 
\nonumber\\ && 
- \frac{640}{9} S_1
-16 \zeta_3
\Biggr]
- \textcolor{blue}{T_F^2}\left[
\frac{64}{3} (\textcolor{blue}{N_F}-2) \zeta_2 + \frac{4}{9} (71 \textcolor{blue}{N_F}+143)\right]
\Biggr\}
\nonumber\\ &&
+\textcolor{blue}{C_F^2 T_F}  
\Biggl[
\frac{P_{30}}{(N-1) N^3 (N+1)^3 (N+2)}
+ (80-128 \ln(2)) 
\zeta_2
+32 \zeta_3
\Biggr],
\nonumber\\ 
\end{eqnarray}
with the polynomials
\begin{eqnarray}
P_{27} &=& 11 N^4+22 N^3-59 N^2-70 N-48,
\\
P_{28} &=& 257 N^6+771 N^5+521 N^4-243 N^3+230 N^2+480 N+144,
\\
P_{29} &=& 1495 N^6+4485 N^5+3927 N^4+379 N^3+3026 N^2+3584 N+768,
\\
P_{30} &=& -13 N^8-52 N^7+76 N^6+282 N^5+129 N^4-614 N^3-320 N^2-256 N-256,
\nonumber\\
\\
P_{31} &=& 5 N^8+20 N^7+12 N^6-10 N^5+75 N^4+254 N^3+188 N^2+112 N+48~.
\end{eqnarray}
Here we have set the masses in both schemes equal symbolically, to obtain a more compact expression.
\subsection{Anomalous Dimension}

\vspace{1mm}
\noindent
As a by-product of the calculation we obtain the corresponding contributions to the
anomalous dimensions from the single pole term $1/\ep$ or the corresponding linear logarithmic 
term, cf. Eq.~(\ref{eqAggQ}),
\begin{eqnarray}
\label{eq:ggg3}
  \hat{\gamma}_{gg}^{(2), \rm T_F^2 C_{F,A}} &=&
-\textcolor{blue}{C_A T_F^2}
\frac{4}{27}
\Biggl\{
\frac{Q_2}{(N-1) N^3 (N+1)^3 (N+2)}
+\frac{4 Q_1}{(N-1) N^2 (N+1)^2 (N+2)} S_1
\Biggr\}
\nonumber\\ &&
+\textcolor{blue}{C_F T_F^2}
\Biggl\{
-\frac{8 Q_3}{27 (N-1) N^4 (N+1)^4 (N+2)}
+\frac{64 P_4}{9 (N-1) N^3 (N+1)^3 (N+2)} S_1
\nonumber\\ &&
+\frac{32}{3} \frac{F}{N(N+1)} [S_1^2 - 3 S_2]
\Biggr\},
\end{eqnarray}
where
\begin{eqnarray}
Q_1 &=& 8 N^6+24 N^5-19 N^4-78 N^3-253 N^2-210 N-96,
\\
Q_2 &=& 87 N^8+348 N^7+848 N^6+1326 N^5+2609 N^4+3414 N^3+2632 N^2
\nonumber\\ &&
+1088 N+192,
\\
Q_3 &=& 33 N^{10}+165 N^9+256 N^8-542 N^7-3287 N^6-8783 N^5-11074 N^4-9624 N^3
\nonumber\\ &&
-5960 N^2-2112 N-288~.
\end{eqnarray}
Eq.~(\ref{eq:ggg3}) confirms previous results in \cite{Vogt:2004mw}
by a first direct diagrammatic calculation, here for massive graphs containing two
fermion lines of equal mass. In Ref.~\cite{Blumlein:2012vq} the anomalous dimension
has been confirmed for 3-loop graphs containing one massless and a massive fermion line.
\section{Conclusions}
\label{sec:5}

\vspace{1mm}
\noindent
The contribution of $O(T_F^2 C_{F,A})$ to the massive operator matrix element $A_{gg,Q}(N)$ at 3-loop order has been 
calculated. It receives contributions from diagrams with two internal massive quark lines of equal mass. The OME can be 
expressed in terms of harmonic sums, supplemented by a single new binomially weighted harmonic sum. The analytic continuation
to $N \in \mathbb{C}$ is given by the recurrence relation of the expressions and the asymptotic representation. The OME has poles 
for $N \in \mathbb{Z}, N \leq 1$. The results have been given for both the on-shell and $\overline{\rm MS}$-scheme for the heavy 
quark mass. In the latter scheme, terms $\propto \zeta_2$ are not present, cf. also Ref.~\cite{BBK09NPB}. As a by-product the 
corresponding contribution to the 3-loop anomalous dimension $\gamma_{gg}$ has been obtained in an independent calculation ab 
initio. The calculation of the diagrams with two massive fermion lines need more special techniques than in the case of a 
single fermion line. Here the use of Mellin-Barnes representations and generating functions based on cyclotomic harmonic 
polylogarithms and S-sums is essential. In some of the diagrams we applied the method the integration by parts method and 
applied differential equations to calculate the associated master integrals. The technologies described can be generalized to 
the case of two different masses.

\vspace{5mm}\noindent
{\bf Acknowledgment.}~
We would like to thank A.~Behring, C. Raab, and F. Wi\ss{}brock for discussions, M.~Steinhauser
for providing the
code {\tt MATAD 3.0}, and  A.~Behring for technical checks of the formulae.
The graphs in the present paper were drawn using {\tt Axodraw} \cite{Vermaseren:1994je}.
This work was supported in part by DFG
Sonderforschungsbereich Transregio 9, Computergest\"utzte Theoretische Teilchenphysik, Studienstiftung des
Deutschen Volkes, the Austrian Science Fund (FWF) grants P20347-N18 and SFB F50 (F5009-N15), the European
Commission through contract PITN-GA-2010-264564 ({LHCPhenoNet}), PITN-GA-2012-316704 ({HIGGSTOOLS}), and the Research Center
{\it Elementary Forces and Mathematical Foundations} (EMG) of J.~Gutenberg University Mainz, the German Research Foundation
(DFG).
\begin{appendix}
\section{Results for the scalar graphs}
\label{app:Results}

\vspace*{1mm}
\noindent
In the following, the results for the scalar prototypes of the graphs
contributing to the $O(T_F^2 C_{F,A})$ part of the operator matrix element
$A_{gg,Q}^{(3)}$ are summarized. These diagrams are much simpler to  calculate than
the corresponding complete diagrams. However, they show the principal structures of the full diagrams 
and share a common calculational scheme. The large amount of numerator terms and their variation, however, increases
the complexity of the QCD diagrams significantly.

All diagrams are normalized such that the factor
\begin{align}
 i a_s^3 S_{\ep}^3 \left(\frac{m^2}{\mu^2}\right)^{\thep - 3} (\Delta.p)^N
\end{align}
is omitted.
The results for the diagrams in Figures~2--8, calculated as explained before,  are given by
\begin{center}
\begin{figure}[H]
\centering
\includegraphics[width=.25\textwidth]{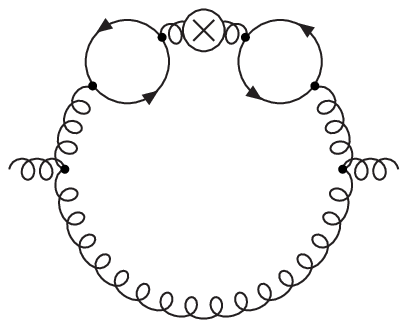}
\refstepcounter{graphnr}
\label{gra:bb1}
\\
\caption[]{\sf Graph \arabic{graphnr}}
\end{figure}
\end{center}
\begin{align}
  \text{Res}_{\arabic{graphnr}}
 ={}& 
  \frac{(-1)^N+1}{2} \Biggl\{
   \frac{2}{105 \ep^2 (N+1)}
  -\frac{1}{\ep} \Biggl[
     \frac{S_{1}}{105 (N+1)} 
    +\frac{57 N+127}{7350 (N+1)^2}
   \Biggr]
\N\\&
  +\frac{1}{420 (N+1)}\left( S_{1}^2 + S_{2} + \zeta_2\right)
  +\frac{57 N+127}{14700 (N+1)^2} S_{1}
  -\frac{75253 N^2+78686 N-84767}{18522000 (N+1)^3}
  \Biggr\}
\comma
\end{align}
\begin{center}
\begin{figure}[H]
\centering
\includegraphics[width=.25\textwidth]{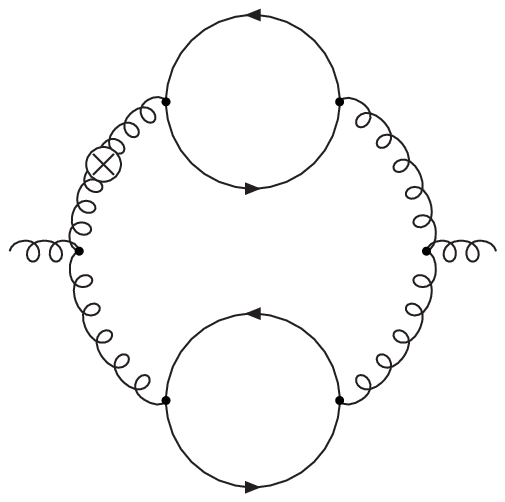}
\refstepcounter{graphnr}
\label{gra:bb2}
\\
\caption[]{\sf Graph \arabic{graphnr}}
\end{figure}
\end{center}
\begin{align}
 \text{Res}_{\arabic{graphnr}}
 ={}& 
  \frac{(-1)^N+1}{2} \Biggl\{
  \frac{1}{105 \ep^2}
 +\frac{1 }{\ep}
  \Biggl[
    \frac{74 N^3-455 N^2+381 N-210}
    {44100 (N-1) N (N+1)}
   -\frac{1}{210} S_{1}
  \Biggr]
\N\\&
 +\frac{8903 N^3+39537 N^2-114440 N+36576}
  {2822400 (N+1) (2 N-3) (2 N-1)} S_{1}
\N\\&
 +\frac{P_{32}}
  {148176000 (N-1)^2 N^2 (N+1)^2 (2 N-3) (2 N-1)}
 +\frac{1}{840} 
  \Bigl( 
   S_{1}^2 + S_{2} + 3 \zeta_2
  \Bigr)
\N\\&
 +\frac{(N-1) N (5 N-6)}
  {1536 (2 N-3) (2 N-1) 4^N}
  \binom{2N}{N}
  \Biggl[
     \sum_{j=1}^N 
     \frac{4^{j} S_{1}(j-1)}
     {\binom{2j}{j} j^2}
   - 7 \zeta_3
   \Biggr]
  \Biggr\}
\comma
\end{align}
\begin{align}
 P_{32} ={}& 
    1795487 N^8
   -7087789 N^7
   +10654130 N^6
   -5797102 N^5
   +6828839 N^4
   -16594069 N^3
\N\\&
   +9651144 N^2
   +902160 N
   -1058400
\comma
\end{align}
\begin{center}
\begin{figure}[H]
\centering
\includegraphics[width=0.25\textwidth]{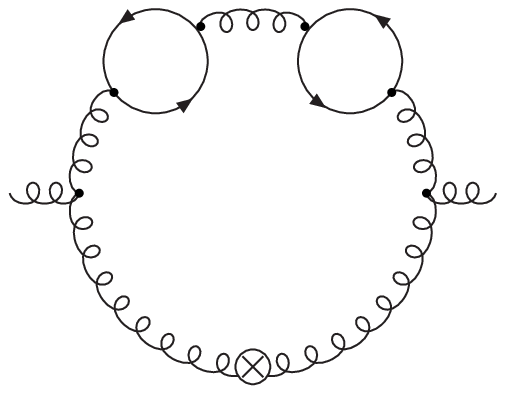}
\refstepcounter{graphnr}
\label{gra:bb3}
\\
\caption[]{\sf Graph \arabic{graphnr}}
\end{figure}
\end{center}
\begin{align}
  \text{Res}_{\arabic{graphnr}}
 ={}&
  \frac{(-1)^N+1}{2} \Biggl\{
  \frac{1}{\ep}
  \frac{1}{105 N (N+1)}
 -\frac{57 N^2+197 N+70}{14700 N^2 (N+1)^2}
  \Biggr\}
\comma
\end{align}
\begin{center}
\begin{figure}[H]
\centering
\includegraphics[width=0.25\textwidth]{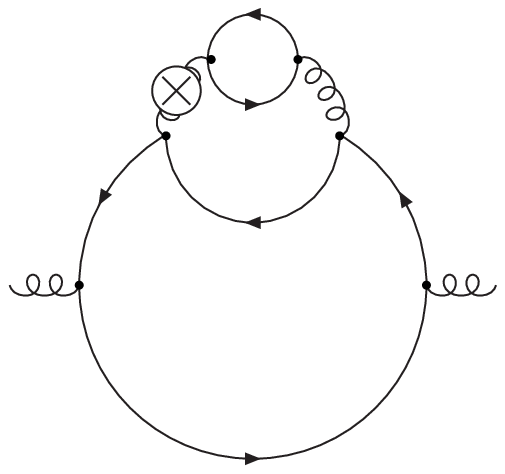}
\refstepcounter{graphnr}
\label{gra:rb2}
\\
\caption[]{\sf Graph \arabic{graphnr}}
\end{figure}
\end{center}
\begin{align}
  \text{Res}_{\arabic{graphnr}}
 ={}&
  \frac{(-1)^N+1}{2} \Biggl\{
 -\frac{1}{\ep}\frac{1}{5 (N-1) N (N+1)^2 (N+2)}
 -\frac{\left(3 N^2-N+56\right)}{192 (N+1)^2 (N+2) (2 N-3) (2 N-1)} S_{1}
\N\\&
 -\frac{(N-3) }{128 (N+1) (2 N-3) (2 N-1) 4^N}
  \binom{2 N}{N}
  \left[
     \sum _{j=1}^N \frac{4^j}{\binom{2 j}{j} j^2} S_{1}(j-1)
    -7 \zeta_3\right]
\N\\&
 -\frac{P_{33}}{7200 (N-1)^2 N^2 (N+1)^3 (N+2) (2 N-3) (2 N-1)}
  \Biggr\}
\comma
\\
 P_{33} ={}& 225 N^7-325 N^6-10398 N^5+6806 N^4+23517 N^3-18721 N^2-1824 N+2160
\comma
\end{align}
\begin{center}
\begin{figure}[H]
\centering
\includegraphics[width=0.25\textwidth]{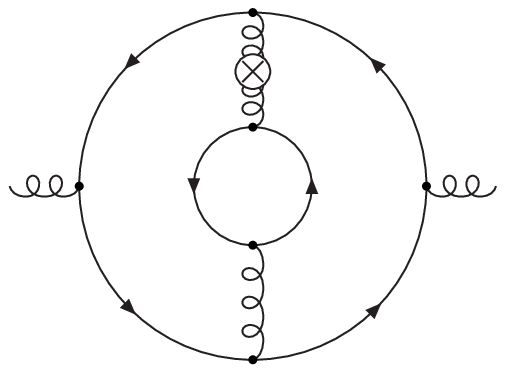}
\refstepcounter{graphnr}
\label{gra:rb}
\\
\caption[]{\sf Graph \arabic{graphnr}}
\end{figure}
\end{center}
\begin{align}
  \text{Res}_{\arabic{graphnr}}
 ={}&
  \frac{(-1)^N+1}{2} 
  \Biggl\{
   -\frac{1}{\ep}\frac{4}{15 (N-1) N (N+1)^2 (N+2)}
\N\\&
   +\frac{N^2-3 N+6}{64 (N+1) (N+2) (2 N-3) (2 N-1) 4^N}
    \binom{2 N}{N} 
    \left[
      \sum _{j=1}^N \frac{4^j}{\binom{2 j}{j} j^2} S_{1}(j-1)
     -7 \zeta_3
    \right]
\N\\&
   +\frac{(N-5) (3 N+8)}{96 (N+1)^2 (N+2) (2 N-3) (2 N-1)} S_{1}
\N\\&
   +\frac{P_{34}}{3600 (N-1)^2 N^2 (N+1)^3 (N+2) (2 N-3) (2 N-1)}
  \Biggr\}
\comma
\end{align}
\begin{align}
 P_{34} = 225 N^7-775 N^6+7702 N^5-4194 N^4-16783 N^3+13129 N^2+1176 N-1440
\comma
\end{align}
\begin{center}
\begin{figure}[H]
\centering
\includegraphics[width=.25\textwidth]{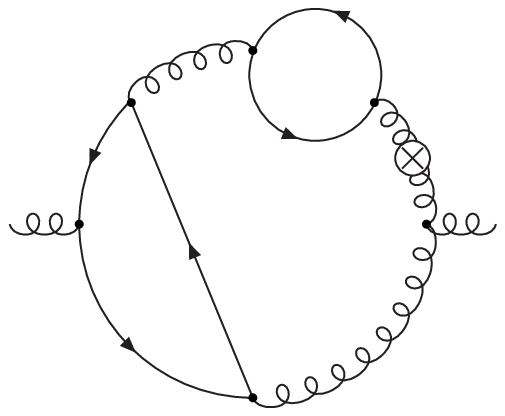}
\refstepcounter{graphnr}
\label{gra:tb}
\\
\caption[]{\sf Graph \arabic{graphnr}}
\end{figure}
\end{center}
\begin{align}
  \text{Res}_{\arabic{graphnr}}
 ={}& 
  \frac{(-1)^N+1}{2} \Biggl\{
   \frac{1}{45 \ep^2 (N+1)}
  -\frac{1}{\ep}
   \Biggl[\frac{S_{1}}{90 (N+1)}+\frac{47 N^3+20 N^2-67 N+40}{1800 (N-1) N (N+1)^2} \Biggr]
\N\\&
  +\frac{ 105 N^3-175 N^2+56 N+96}{13440 (N+1)^2 (2 N-3) (2 N-1) 4^N} \binom{2 N}{N} 
   \left[
     \sum _{j=1}^N \frac{4^j S_{1}(j-1)}{\binom{2 j}{j} j^2}
    -7 \zeta_3\right]
\N\\&
  +\frac{\left(5264 N^3-2409 N^2-12770 N+3528\right) S_{1}}{100800 (N+1)^2 (2 N-3) (2 N-1)}
  +\frac{S_{1}^2+S_{2}+3 \zeta_2}{360 (N+1)}
\N\\&
  +\frac{S_{3}-S_{2,1}+7 \zeta_3}{420 (N+1)}
  +\frac{P_{35}}{2268000 (N-1)^2 N^2 (N+1)^3 (2 N-3) (2 N-1)}
  \Biggr\}
\comma
\end{align}
\begin{align}
 P_{35} ={}& -257476 N^8+682667 N^7-144175 N^6-586654 N^5+615368 N^4-948403 N^3+592683 N^2
\N\\&
 +71190 N-75600
\comma
\end{align}
\begin{center}
\begin{figure}[H]
\centering
\includegraphics[width=.25\textwidth]{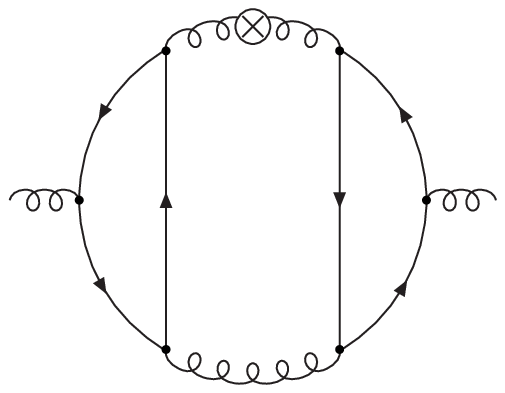}
\refstepcounter{graphnr}
\label{gra:tt}
\\
\caption[]{\sf Graph \arabic{graphnr}}
\end{figure}
\end{center}
\begin{align}
  \text{Res}_{\arabic{graphnr}}
 ={}& 
  \frac{(-1)^N+1}{2} \Biggl\{
  \frac{27 N^2+49 N+38}
   {2880 (N+1)^2 (N+2) 4^{N}} 
  \binom{2 N}{N}
  \Biggl[
    \sum _{j=1}^N \frac{4^{j}}{\binom{2 j}{j} j^2} S_{1}(j-1)
   -7 \zeta_3
  \Biggr]
\N\\&
 +\frac{1}{90 (N+1)} 
   \Bigl[S_{3} - S_{2,1} + 7 \zeta_3\Bigr]
 +\frac{1}{90 N (N+1)^2 (N+2)} \Bigl[ S_{2} - S_{1}^2 \Bigr]
\N\\&
 +\frac{60 N^2+191 N+120}{1440 (N+1)^2 (N+2)} S_{1}
 -\frac{81 N^3+194 N^2+83 N+60}{720 N (N+1)^2 (N+2)}
\N\\&
 -\frac{1}{\ep} \frac{1}{12 (N+1)}
  \Biggr\}
\comma
\end{align}
\begin{center}
\begin{figure}[H]
\centering
\includegraphics[width=0.25\textwidth]{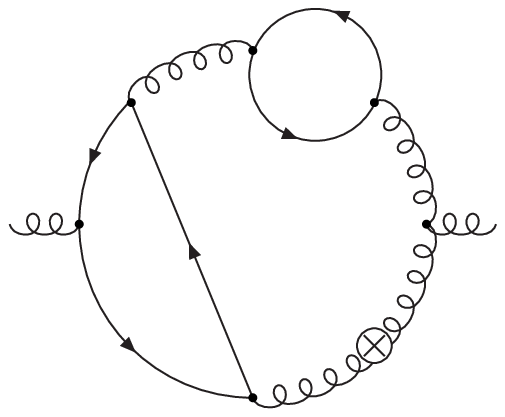}
\refstepcounter{graphnr}
\label{gra:tb2}
\\
\caption[]{\sf Graph \arabic{graphnr}}
\end{figure}
\end{center}
\begin{align}
  \text{Res}_{\arabic{graphnr}}
 ={}&
  \frac{(-1)^N+1}{2} \Biggl\{
    \frac{1}{\ep^2}\frac{N+2}{45 (N+1)}
   +\frac{1}{\ep}
    \Bigg[
      \frac{(N-4) \big(8 N^2+11 N-5\big)}{1800 N (N+1)^2}
     -\frac{N+2}{90 (N+1)} S_1
    \Bigg]
\N\\&
   +\frac{25 N^3+81 N^2+72 N+32}{13440 (N+1)^2 4^N} 
    \binom{2 N}{N}
    \Bigg[
      \sum_{j=1}^N \frac{4^j}{j^2 \binom{2 j}{j}} S_1(j-1)
     -7 \zeta_3
    \Bigg]
\N\\&
   +\frac{151 N^2+1678 N+2072}{100800 (N+1)^2} S_1
   +\frac{7 N^3+21 N^2+14 N-3}{2520 N (N+1)^2} S_1^2
\N\\&
   +\frac{7 N^3+21 N^2+14 N+3}{2520 N (N+1)^2} S_2
   +\frac{N+2}{120 (N+1)} \zeta_2
\N\\&
   +\frac{16091 N^5+37499 N^4+46885 N^3-4133 N^2-67410 N-12600}{2268000 N^2 (N+1)^3}
  \Biggr\}~.
\end{align}

\noindent
In some of the graphs, the denominator structure show evanescent poles at $N = 1/2, 3/2, 5/2$. The expansion
of the whole function around these values shows continuity.
\end{appendix}

\newpage


\begin{thebibliography}{99}
%
\bibitem{Bethke:2011tr}
  S.~Bethke et al.
  {\sf Workshop on Precision Measurements of $\alpha_s$}, 
  arXiv:1110.0016 [hep-ph].
%
\bibitem{PDF}
  S.~Alekhin, J.~Bl\"umlein and S.~Moch,
  Phys.\ Rev.\ D {\bf 89} (2014) 054028
  [arXiv:1310.3059 [hep-ph]];\\
  E.~Perez and E.~Rizvi,
  Rep.\ Prog.\ Phys.\  {\bf 76} (2013) 046201
  [arXiv:1208.1178 [hep-ex]];\\
  J.~Bl\"umlein,
  Prog.\ Part.\ Nucl.\ Phys.\  {\bf 69} (2013) 28
  [arXiv:1208.6087 [hep-ph]].
%
\bibitem{HEAV2}
  E.~Laenen, S.~Riemersma, J.~Smith and W.~L.~van Neerven,
  {Nucl.\ Phys.}\ B {\bf 392} (1993) 162;
229;\\
  S.~Riemersma, J.~Smith and W.~L.~van Neerven,
  {Phys.\ Lett.}\ B {\bf 347} (1995) 143
  [hep-ph/9411431].
%
\bibitem{Alekhin:2003ev}
  S.~I.~Alekhin and J.~Bl\"umlein,
  Phys.\ Lett.\ B {\bf 594} (2004) 299
  [hep-ph/0404034].
%
\bibitem{BMSN96}
  M.~Buza, Y.~Matiounine, J.~Smith, R.~Migneron and W.~L.~van Neerven,
  Nucl.\ Phys.\ B {\bf 472} (1996) 611
  [hep-ph/9601302].
%
\bibitem{BBK09NPB}
  I.~Bierenbaum, J.~Bl\"umlein and S.~Klein,
  {Nucl.\ Phys.}\ B {\bf 820} (2009) 417
  [arXiv:0904.3563 [hep-ph]].
%
\bibitem{MVV2005}
  J.~A.~M.~Vermaseren, A.~Vogt and S.~Moch,
  Nucl.\ Phys.\ B {\bf 724} (2005) 3
  [hep-ph/0504242].
%
\bibitem{BBK07NPB}
  I.~Bierenbaum, J.~Bl\"umlein and S.~Klein,
  {Nucl.\ Phys.}\ B {\bf 780} (2007) 40
  [hep-ph/0703285].
%
\bibitem{BMSN98}
  M.~Buza, Y.~Matiounine, J.~Smith and W.~L.~van Neerven,
  {Eur.\ Phys.\ J.}\ C {\bf 1} (1998) 301
  [hep-ph/9612398].
%
\bibitem{BBK09PLB}
  I.~Bierenbaum, J.~Bl\"umlein and S.~Klein,
  {Phys.\ Lett.}\ B {\bf 672} (2009) 401
  [arXiv:0901.0669 [hep-ph]].
%
\bibitem{Buza:1996xr}
  M.~Buza, Y.~Matiounine, J.~Smith and W.~L.~van Neerven,
  {Nucl.\ Phys.}\ B {\bf 485} (1997) 420
  [hep-ph/9608342].
%
\bibitem{Bierenbaum:2007pn}
  I.~Bierenbaum, J.~Bl\"umlein and S.~Klein,
  {PoS (ACAT)} 070, arXiv:0706.2738 [hep-ph].
%
\bibitem{BBKS08NPB}
  I.~Bierenbaum, J.~Bl\"umlein, S.~Klein and C.~Schneider,
  {Nucl.\ Phys.}\ B {\bf 803} (2008) 1
  [arXiv:0803.0273 [hep-ph]].
%
\bibitem{Blumlein:2009rg}
  J.~Bl\"umlein, S.~Klein and B.~T\"odtli,
  {Phys.\ Rev.}\ D {\bf 80} (2009) 094010
  [arXiv:0909.1547 [hep-ph]].
%
\bibitem{CC}
  T.~Gottschalk,
  Phys.\ Rev.\  D {\bf 23} (1981) 56;\\
  M.~Gl\"uck, S.~Kretzer and E.~Reya,
  Phys.\ Lett.\  B {\bf 380} (1996) 171
  [Erratum-ibid.\  B {\bf 405} (1997) 391]
  [arXiv:hep-ph/9603304];\\
  J.~Bl\"umlein, A.~Hasselhuhn, P.~Kovacikova and S.~Moch,
  {Phys.\ Lett.}\ B {\bf 700} (2011) 294
  [arXiv:1104.3449 [hep-ph]];\\
  M.~Buza and W.~L.~van Neerven,
  {Nucl.\ Phys.}\ B {\bf 500} (1997) 301
  [hep-ph/9702242];\\
  J.~Bl\"umlein, A.~Hasselhuhn and T.~Pfoh,
  Nucl.\ Phys.\ B {\bf 881} (2014) 1
  [arXiv:1401.4352 [hep-ph]].
%
\bibitem{LOGS}
  A.~Behring, I.~Bierenbaum, J.~Bl\"umlein, A.~De Freitas, S.~Klein and F.~Wi\ss{}brock,
  arXiv:1403.6356 [hep-ph].
%
\bibitem{Blumlein:2006mh}
  J.~Bl\"umlein, A.~De Freitas, W.~L.~van Neerven and S.~Klein,
  Nucl.\ Phys.\ B {\bf 755} (2006) 272
  [hep-ph/0608024].
%
\bibitem{ABKSW11NPB}
  J.~Ablinger, J.~Bl\"umlein, S.~Klein, C.~Schneider and F.~Wi\ss{}brock,
  Nucl.\ Phys.\ B {\bf 844} (2011) 26
  [arXiv:1008.3347 [hep-ph]].
%
\bibitem{Blumlein:2012vq}
  J.~Bl\"umlein, A.~Hasselhuhn, S.~Klein and C.~Schneider,
  Nucl.\ Phys.\ B {\bf 866} (2013) 196
  [arXiv:1205.4184 [hep-ph]].
%
\bibitem{ABHKSW12}
  J.~Ablinger, J.~Bl\"umlein, A.~Hasselhuhn, S.~Klein, C.~Schneider and F.~Wi\ss{}brock,
  Nucl.\ Phys.\ B {\bf 864} (2012) 52
  [arXiv:1206.2252 [hep-ph]].
%
\bibitem{Ablinger:2014yaa}
  J.~Ablinger, J.~Bl\"umlein, C.~Raab, C.~Schneider and F.~Wi\ss{}brock,
  arXiv:1403.1137 [hep-ph].
%
\bibitem{Ablinger:2011pb}
  J.~Ablinger, J.~Bl\"umlein, S.~Klein, C.~Schneider and F.~Wi\ss{}brock,
  arXiv:1106.5937 [hep-ph].
%
\bibitem{Ablinger:2012qj}
  J.~Ablinger, J.~Bl\"umlein, A.~Hasselhuhn, S.~Klein, C.~Schneider and 
  F.~Wi\ss{}brock,
  arXiv:1202.2700 [hep-ph].
%
\bibitem{Ablinger:2014lka}
  J.~Ablinger, J.~Bl\"umlein, A.~De Freitas, A.~Hasselhuhn, A.~von Manteuffel, M.~Round, C.~Schneider and F.~Wi\ss{}brock,
  Nucl.\ Phys.\ B {\bf 882} (2014) 263
  [arXiv:1402.0359 [hep-ph]].
%
\bibitem{NSPS}
J.~Ablinger et al., DESY 13--210, DESY 13--232.
%
\bibitem{Blumlein:2011mi}
  J.~Bl\"umlein, A.~De Freitas and W.~van Neerven,
  {Nucl.\ Phys.}\  B {\bf 855} (2012) 508
  [arXiv:1107.4638 [hep-ph]].
%
\bibitem{HSUM}
  J.~A.~M.~Vermaseren,
  {Int.\ J.\ Mod.\ Phys.}\ A {\bf 14} (1999) 2037
  [hep-ph/9806280];\\
  J.~Bl\"umlein and S.~Kurth,
  {Phys.\ Rev.}\ D {\bf 60} (1999) 014018
  [hep-ph/9810241].
%
\bibitem{ABRS14}
J. Ablinger, J. Bl\"umlein, C. Raab and C. Schneider, DESY 14--021.
%
\bibitem{BINO}
  M.~Y.~Kalmykov and O.~Veretin,
  Phys.\ Lett.\ B {\bf 483} (2000) 315
  [hep-th/0004010];\\
  A.~I.~Davydychev and M.~Y.~.Kalmykov,
  Nucl.\ Phys.\ B {\bf 699} (2004) 3
  [hep-th/0303162];\\
  S.~Weinzierl,
  J.\ Math.\ Phys.\  {\bf 45} (2004) 2656
  [hep-ph/0402131];\\
  M.~Y.~.Kalmykov, B.~F.~L.~Ward and S.~A.~Yost,
  JHEP {\bf 0710} (2007) 048
  [arXiv:0707.3654 [hep-th]].
%
\bibitem{Fleischer:1998nb}
  J.~Fleischer, A.~V.~Kotikov and O.~L.~Veretin,
  Nucl.\ Phys.\ B {\bf 547} (1999) 343
  [hep-ph/9808242].
%
\bibitem{SIGMA}
C. Schneider, { J. Symbolic Comput.} {\bf 43} (2008) 611, \newblock [arXiv:0808.2543v1]; 
{Ann. Comb.} {\bf 9} (2005) 75; {J. Differ. Equations Appl. }{\bf 11} (2005) 799; 
{Ann. Comb. } {\bf 14} (4) (2010), [arXiv:0808.2596]; Proceedings of the Workshop 
{\sf Motives, Quantum Field Theory, and Pseudodifferential Operators}, held at the 
Clay 
Mathematics Institute, Boston University, June 2--13, 2008, Clay Mathematics 
Proceedings {\bf 12} (2010) pp.~285
Eds. A.~Carey, D.~Ellwood, S.~Paycha, S.~Rosenberg;
{S\'em.~Lothar. Combin.} {\bf 56} (2007) 1, Article B56b,  Habilitationsschrift JKU Linz 
(2007) and references therein;\\
  J.~Ablinger, J.~Bl\"umlein, S.~Klein, C.~Schneider,
  {Nucl.\ Phys.\ (Proc.\ Suppl.)}\  {\bf 205-206 } (2010)  110
  [arXiv:1006.4797 [math-ph]];\\
C.~Schneider, in~: Lecture Notes in Computer Science (LNCS) eds. J.~Gutierrez, J.~Schicho, M.~Weimann, in press, 
arXiv:1307.7887[cs.SC] (2013).
%
\bibitem{Harmonicsums}
  J.~Ablinger,
  {\sf A Computer Algebra Toolbox for Harmonic Sums Related to Particle Physics}, Master's Thesis, JKU Linz,
  arXiv:1011.1176 [math-ph];
  {\sf Computer Algebra Algorithms for Special Functions in Particle Physics}, PhD Thesis, JKU Linz,
  arXiv:1305.0687 [math-ph].
%
\bibitem{Ablinger:2011te}
  J.~Ablinger, J.~Bl\"umlein and C.~Schneider,
  J.\ Math.\ Phys.\  {\bf 52} (2011) 102301
  [arXiv:1105.6063 [math-ph]] and in preparation.
%
\bibitem{Ablinger:2013cf}
  J.~Ablinger, J.~Bl\"umlein and C.~Schneider,
  J.\ Math.\ Phys.\  {\bf 54} (2013) 082301
  [arXiv:1302.0378 [math-ph]].
%
\bibitem{EMSSP}
  J.~Ablinger, J.~Bl\"umlein, S.~Klein and C.~Schneider,
  Nucl.\ Phys.\ Proc.\ Suppl.\  {\bf 205-206} (2010) 110
  [arXiv:1006.4797 [math-ph]];\\
  J.~Bl\"umlein, A.~Hasselhuhn and C.~Schneider,
  PoS RADCOR {\bf 2011} (2011) 032
  [arXiv:1202.4303 [math-ph]];\\
  C.~Schneider,
  arXiv:1310.0160 [cs.SC], {Proc. of ACAT 2013} in press.
%
\bibitem{RHOSUM}
C.~Schneider, Advances in Applied Math. {\bf 34} (4) (2005) 740;\\
  J.~Ablinger, J.~Bl\"umlein, M.~Round and C.~Schneider,
  PoS LL {\bf 2012} (2012) 050
   [PoS LL {\bf 2012} (2012) 050]
  [arXiv:1210.1685 [cs.SC]];\\
M.~Round et al., in preparation.
%
\bibitem{IBP}
J. Lagrange, {\sf Nouvelles recherches sur la nature et la propagation
du son}, Miscellanea Taurinensis, t. II, 1760-61; Oeuvres t. I, p. 263;\\
C.F. Gauss, {Theoria attractionis corporum sphaeroidicorum ellipticorum
homogeneorum methodo novo tractate}, Commentationes societas scientiarum
Gottingensis recentiores, Vol III, 1813, Werke Bd. {\bf V} pp. 5-7;\\
G. Green, {\sf Essay on the Mathematical Theory of Electricity and
Magnetism}, Nottingham, 1828 [Green Papers, pp. 1-115];\\
M. Ostrogradski, Mem. Ac. Sci. St. Peters., {\bf 6}, (1831) 39;\\
  K.~G.~Chetyrkin, A.~L.~Kataev and F.~V.~Tkachov,
  Nucl.\ Phys.\  B {\bf 174} (1980) 345.
%
\bibitem{Vogt:2004mw}
  A.~Vogt, S.~Moch and J.~A.~M.~Vermaseren,
  Nucl.\ Phys.\ B {\bf 691} (2004) 129
  [hep-ph/0404111].
%
\bibitem{Bierenbaum:2010jp}
  I.~Bierenbaum, J.~Bl\"umlein and S.~Klein,
  {PoS} {\bf DIS2010} (2010) 148
  [arXiv:1008.0792 [hep-ph]].
%
\bibitem{Steinhauser:2000ry}
  M.~Steinhauser,
  Comput.\ Phys.\ Commun.\  {\bf 134} (2001) 335
  [hep-ph/0009029].
%
\bibitem{Nogueira:1991ex}
  P.~Nogueira, 
  J.\ Comput.\ Phys.\  {\bf 105} (1993) 279.
%
\bibitem{Klein:2009ig}
  S.~W.~G.~Klein,
  {\sf Mellin Moments of Heavy Flavor Contributions to $F_2(x,Q^2)$ at NNLO},
  arXiv:0910.3101 [hep-ph].
%
\bibitem{Hasselhuhn:2013swa}
  A.~Hasselhuhn,
  {\sf 3-Loop Contributions to Heavy Flavor Wilson Coefficients of Neutral and Charged Current DIS},
  DESY-THESIS-2013-050.
%
\bibitem{vanRitbergen:1998pn}
  T.~van Ritbergen, A.~N.~Schellekens and J.~A.~M.~Vermaseren,
  Int.\ J.\ Mod.\ Phys.\  A {\bf 14}  (1999) 41
  [arXiv:hep-ph/9802376].
%
\bibitem{Hamberg91}
R.~Hamberg,
{\sf {Second order gluonic contributions to physical quantities}},
PhD thesis, Leiden University, 1991.
%
\bibitem{Bierenbaum:2007dm}
  I.~Bierenbaum, J.~Bl\"umlein and S.~Klein,
  Phys.\ Lett.\ B {\bf 648} (2007) 195
  [hep-ph/0702265].
%
\bibitem{Bierenbaum:2007qe}
  I.~Bierenbaum, J.~Bl\"umlein and S.~Klein,
  Nucl.\ Phys.\ B {\bf 780} (2007) 40
  [hep-ph/0703285].
%
\bibitem{Mellin1895}
H.~Mellin,
{Acta Societatis Scientiarum Fennicae}, {\bf XX.}(7) (1895)
  1;
Math. Ann. {\bf 68} (1910) 305.
%
\bibitem{Barnes1908}
E.W. Barnes, Proc. Lond. Math. Soc. (2) {\bf 6} (1908) 141; Quart. J.
Math. {\bf 41} (1910) 136.
%
\bibitem{WWT}
E.T. Whittaker and G.N. Watson, {\sf A Course of Modern Analysis}, (Cambridge
University
Press, Cambridge, 1927; reprinted 1996);\\
E.C. Titchmarsh, {\sf Introduction to the Theory of Fourier Integrals},
(Oxford,
Calendron Press,
1937; 2nd Edition 1948). 
%
\bibitem{Smirnov2006}
V.~A. Smirnov, {\sf {Feynman Integral Calculus}},
(Springer, Berlin, 2006).
%
\bibitem{HYP}
W.N. Bailey, {\sf Generalized Hypergeometric Series}, (Cambridge University
Press,  Cambridge, 1935);\\
A. Erd\'{e}lyi {\it et al.}, {\sf H.~Bateman Manuscript Project, Higher Transcendental Functions},
Vol.~{\bf I}, (McGraw--Hill, New Your, 1953);\\
P. Appell and J. Kamp\'{e} de F\'{e}riet, {\sf Fonctions
Hyperg\'{e}om\'{e}triques et Hypersp\'{e}riques, Polynomes D' Hermite},
(Gauthier-Villars, Paris, 1926);\\
P. Appell, {\sf Les Fonctions Hyperg\'{e}om\'{e}triques de Plusieur
Variables}, (Gauthier-Villars, Paris, 1925);\\
J. Kamp\'{e} de F\'{e}riet, {\sf La fonction
hyperg\'{e}om\'{e}trique},(Gauthier-Villars, Paris, 1937);\\
H. Exton, {\sf Multiple Hypergeometric Functions and Applications},
(Ellis Horwood, Chichester, 1976).\\
H. Exton, {\sf Handbook of Hypergeometric Integrals},
(Ellis Horwood, Chichester, 1978).\\
H.M. Srivastava and P.W. Karlsson, {\sf Multiple Gaussian Hypergeometric
Series}, (Ellis Horwood, Chicester, 1985).
%
\bibitem{SLATER}
L.J. Slater, {\sf Generalized Hypergeometric Functions}, (Cambridge University
Press, Cambridge, 1966).
%
\bibitem{Tausk:1999vh}
  J.~B.~Tausk,
  Phys.\ Lett.\ B {\bf 469} (1999) 225
  [hep-ph/9909506].
%
\bibitem{Gluza:2007rt}
  J.~Gluza, K.~Kajda and T.~Riemann,
  Comput.\ Phys.\ Commun.\  {\bf 177} (2007) 879
  [arXiv:0704.2423 [hep-ph]].
%
\bibitem{Czakon:2005rk}
  M.~Czakon,
  Comput.\ Phys.\ Commun.\  {\bf 175} (2006) 559
  [hep-ph/0511200].
%
\bibitem{KOSOW}
D.~Kosower,
{\tt https://www.hepforge.org/downloads/mbtools/barnesroutines-1.0.tar.gz}
%
\bibitem{Moch:2001zr}
  S.~Moch, P.~Uwer and S.~Weinzierl,
  J.\ Math.\ Phys.\  {\bf 43} (2002) 3363
  [hep-ph/0110083].
%
\bibitem{Gehrmann:2001jv}
  T.~Gehrmann and E.~Remiddi,
  Comput.\ Phys.\ Commun.\  {\bf 144} (2002) 200
  [hep-ph/0111255].
%
\bibitem{Brown:2008um}
  F.~Brown,
  Commun.\ Math.\ Phys.\  {\bf 287} (2009) 925
  [arXiv:0804.1660 [math.AG]].
%
\bibitem{Remiddi:1999ew}
  E.~Remiddi and J.~A.~M.~Vermaseren,
  Int.\ J.\ Mod.\ Phys.\ A {\bf 15} (2000) 725
  [hep-ph/9905237].
%
\bibitem{Blumlein:2003gb}
  J.~Bl\"umlein,
  Comput.\ Phys.\ Commun.\  {\bf 159} (2004) 19
  [hep-ph/0311046].
%
\bibitem{vonManteuffel:2013vja}
  A.~von Manteuffel, R.~M.~Schabinger and H.~X.~Zhu,
  JHEP {\bf 1403} (2014) 139
  [arXiv:1309.3560 [hep-ph]].
%
\bibitem{FORM}
  J.~A.~M.~Vermaseren,
  math-ph/0010025;\\
  M.~Tentyukov and J.~A.~M.~Vermaseren,
  Comput.\ Phys.\ Commun.\  {\bf 181} (2010) 1419
  [hep-ph/0702279].
%
\bibitem{vonManteuffel:2012np}
  A.~von Manteuffel and C.~Studerus,
  arXiv:1201.4330 [hep-ph];\\
  C.~Studerus,
  Comput.\ Phys.\ Commun.\  {\bf 181} (2010) 1293
  [arXiv:0912.2546 [physics.comp-ph]].
%
\bibitem{FERMAT}
R.H.~Lewis, Computer Algebra System {\tt Fermat}, {\tt http://home.bway.net/lewis.}
%
\bibitem{Bauer:2000cp}
  C.~W.~Bauer, A.~Frink and R.~Kreckel,
  cs/0004015 [cs-sc].
%
\bibitem{Laporta:2001dd}
  S.~Laporta,
  Int.\ J.\ Mod.\ Phys.\ A {\bf 15} (2000) 5087
  [hep-ph/0102033].
%
\bibitem{DEQ}
  A.~V.~Kotikov,
  Phys.\ Lett.\ B {\bf 254} (1991) 158;\\
  M.~Caffo, H.~Czyz, S.~Laporta and E.~Remiddi,
  Acta Phys.\ Polon.\ B {\bf 29} (1998) 2627
  [hep-th/9807119];\\
  Nuovo Cim.\ A {\bf 111} (1998) 365
  [hep-th/9805118];\\
  T.~Gehrmann and E.~Remiddi,
  Nucl.\ Phys.\ B {\bf 580} (2000) 485
  [hep-ph/9912329];\\
  M.~Caffo, H.~Czyz and E.~Remiddi,
  Nucl.\ Phys.\ B {\bf 634} (2002) 309
  [hep-ph/0203256].
%
\bibitem{Gerhold:02}
S.~Gerhold,
{\sf Uncoupling systems of linear ore operator equations},
\newblock Master's thesis, RISC, J.~Kepler University, Linz, 2002.
%
\bibitem{Blumlein:2010zv}
  J.~Bl\"umlein, S.~Klein, C.~Schneider and F.~Stan,
  J. Symbolic Comput. 47 (2012) 1267 
  [arXiv:1011.2656 [cs.SC]].
%
\bibitem{GUESS}
M.~Kauers, {\sf Guessing Handbook}, Technical Report RISC 09-07 (2009), JKU Linz.
%
\bibitem{ANCONT1}
  J.~Bl\"umlein,
  Comput.\ Phys.\ Commun.\  {\bf 180} (2009) 2218
  [arXiv:0901.3106 [hep-ph]];
%
\bibitem{ANCONT}
  J.~Bl\"umlein,
  Comput.\ Phys.\ Commun.\  {\bf 133} (2000) 76
  [hep-ph/0003100];
  in : Proceedings of the Workshop
{\sf Motives, Quantum Field Theory, and Pseudodifferential Operators}, held at the 
Clay
Mathematics Institute, Boston University, June 2--13, 2008, Clay Mathematics
Proceedings {\bf 12} (2010) pp.~167,
~Eds. A.~Carey, D.~Ellwood, S.~Paycha, S.~Rosenberg,
  arXiv:0901.0837 [math-ph];\\
  A.~V.~Kotikov and V.~N.~Velizhanin,
  hep-ph/0501274;\\
  J.~Bl\"umlein and S.~-O.~Moch,
  Phys.\ Lett.\ B {\bf 614} (2005) 53
  [hep-ph/0503188].
%
\bibitem{Vermaseren:1994je}
  J.~A.~M.~Vermaseren,
  Comput.\ Phys.\ Commun.\  {\bf 83} (1994) 45.
\end{thebibliography}
\end{document}